\documentclass[3p,number,sort&compress,times,12pt]{elsarticle}
\usepackage{dcolumn}% Align table columns on decimal point
\usepackage{bm}% bold math
\usepackage{mathrsfs}
\usepackage{pstricks}
\usepackage{color}
\usepackage{slashed}
\usepackage{amsmath}
\usepackage{epsfig}
\usepackage{amsfonts}
\usepackage{amssymb}
\def\beq{\begin{equation}}
\def\eeq{\end{equation}}
\def\bea{\begin{eqnarray}}
\def\eea{\end{eqnarray}}
\def\nn{\nonumber}

\def\Re{\textrm{Re}}
\def\Im{\textrm{Im}}

\def\x{\mathbf{x}}

\def\Res{\textrm{Res}}
\journal{Annals of Physics}

\begin{document}

\begin{frontmatter}

\title{Finite-time response function of uniformly accelerated entangled atoms}
\author[nfs]{C. D. Rodr\'iguez-Camargo}
\ead{christian@cbpf.br}
\address[nfs]{Centro Brasileiro de Pesquisas F\'isicas, 222990-180 Rio de Janeiro, RJ, Brazil}
%\address[cdrc]{Centro Brasileiro de Pesquisas F\'isicas, 222990-180 Rio de Janeiro, RJ, Brazil}
%
\author[nfs,gsm]{G. Menezes\corref{cor1}}
\ead{gabrielmenezes@ufrrj.br}
\address[gsm]{Grupo de F\'isica Te\'orica e Matem\'atica F\'isica, Departamento de F\'isica, 
\\ Universidade Federal Rural do Rio de Janeiro, 23897--000 Serop\'edica, RJ, Brazil}
\author[nfs]{N. F. Svaiter}
\ead{nfuxsvai@cbpf.br}
\cortext[cor1]{Corresponding author}
%
%%%%%%%%%%%%%%%%%%%%%%%%%%%%%%%%%%%%%%%%%%
\begin{abstract}
We examine the transition probability from the ground state to a final entangled state of a system of uniformly accelerated two-level atoms weakly coupled with a massless scalar field in Minkowski vacuum. Using time-dependent perturbation theory we evaluate the finite-time response function and we identify the mutual influence of atoms via the quantum field as a coherence agent in each response function terms. The associated thermal spectrum perceived by the atoms is found for a finite time interval. By considering modifications of specific parameters of our setup, we also analyze how the transition probabilities are affected by the smoothness of the switching of the atom-field coupling. In addition, we study the mean life of the symmetric maximally entangled state for different accelerations. Our calculations reveal that smooth switching is more efficient than sudden switching concerning the reduction of the decay of the entangled state. The possible relevance of our results is discussed. 
\end{abstract}
%%%%%%%%%%%%%%%%%%%%%%%%%%%%%%%%%%%%%%%%%%

\begin{keyword}
Quantum Entanglement \sep Unruh Effect \sep Quantum Field Theory
\end{keyword}

\end{frontmatter}

\section{Introduction}

A detector moving with a constant proper acceleration $a$ will perceive the Min\-kows\-ki vacuum state of a quantum field as a thermal bath with the following temperature: $T=\hbar a /(2\pi ck_{B})$, where $\hbar, k_{B},c$ are the reduced Planck's and Boltzmann's constants and the speed of light, respectively~\cite{davies1,unruh1} . Using perturbation theory in first-order approximation it can be shown that the transition rates of the Unruh-DeWitt detector~\cite{dw1} interacting with a scalar field in the Minkowski vacuum is given by the Fourier transform of the positive frequency Wightman function evaluated on the world line of the detector~\cite{sciama1}. In the case of an uniformly accelerated detector, it is found that, if it is initially prepared in its ground state, it can be excited by the thermal radiation perceived by it~\cite{matsas1}.

Quantum entanglement is considered as one of the key features of quantum information processing.  Several sources of entangled quantum systems have been discussed in literature, for instance, in solid-state physics, quantum optics, and also atoms in cavity quantum electrodynamics~\cite{zeilin1}. Some examples of generation of entangled systems of two-level atoms interacting with a bosonic field can be found in Refs.~\cite{knight1, bash1}. Aside from production of those entangled systems, quantum-information processing requires the presence of a strong coherent coupling between the entities of the system. Nevertheless, under realistic experimental conditions, entanglement is degraded through uncontrolled coupling to environment~\cite{vivi1}.

In recent years the field of relativistic quantum information has emerged as an important research topic and is generating increasing interest within the scientific community. The mutual influence of atoms through their interaction with quantum fields is an important stimulating issue in order to analyze decoherence properties~\cite{hu3,hu2,hu1}. Moreover, an important aspect that has been witnessing a vigorous scrutiny in the recent literature is the possibility of entanglement harvesting. This is the phenomenon in which atoms are able to extract entanglement from the vacuum state of a quantum field. In other words, since quantum vacuum fluctuations are entangled (regarding the state space of local observables), they can act as a source of entanglement for atoms which are coupled with such a quantum field. This is particularly interesting for the case of uniformly accelerated atoms, since it is well known that the Minkowski vacuum state can be expressed as an entangled state between the left and right Rindler wedges when formulated on the Rindler vacuum~\cite{Unruh:84}. Some works studying relativistic quantum entanglement in different setups are given by Refs.~\cite{Valentini:91,ivet1,ivet2,hu1,cliche2,reznik1,Reznick:05,Steeg:09,cliche1,Martinez:12,Salton:15,martin2}. In turn, for a wide set of results in this area and special issues of performing satellite experiments, we refer the reader the Ref.~\cite{rideout1}. Many of such works demonstrate that entanglement is an observer-dependent quantity. In addition, recent works point towards the importance of considering explicitly the contributions of vacuum fluctuations and radiation reaction in the radiative processes of entangled atoms~\cite{gabriel1,gabriel3}. Within such a context, we also refer the reader the Refs.~\cite{sch,kerr}. 

In the present work we wish to address different issues in comparison with the aforementioned papers. Here we are interested in studying the transition probability to entangled states for uniformly accelerated two-level atoms due to the vacuum fluctuations of a quantum scalar field when the atoms are initially prepared in the ground state. We are not attempting to present a full analysis concerning the entanglement contained in the final state of the atoms since this would require more general tools, such as the procedure proposed by Reznik in Ref.~\cite{reznik1}, or even the master equation approach, with a suitable characterization of an entanglement monotone. Instead, given that the uniformly accelerated atomic system is initially prepared in the ground state, we formulate a simple question on the probability that the atomic system can be found in a specific entangled state in a later time due to its coupling to a quantum field. For this investigation we employ a straightforward calculation within a quantum mechanical time-dependent perturbation theory analysis. Furthermore we investigate in detail the transition probability under different description perspectives, encoded in different choices for the coupling between atoms and fields. We remark that this simplified situation contains all the important ingredients to study radiative processes involving entangled states. On the other hand, an important issue that naturally arises in this context is whether these recently formed entangled states could persist for long time intervals. In order to pursue an answer to this question, one may investigate the mean life of the entangled states, which is one of the topics of the present work. In this way we propose to afford an answer as broad as possible for the query of how radiative processes of entangled states are impacted by effects of acceleration in a perturbative framework.

The aim of this paper is to study radiative processes of a pair of uniformly accelerated two-level atoms interacting with a quantum massless scalar field. We evaluate the transition rates within time-dependent perturbation theory in a finite time interval. Refs.~\cite{nami1, nami2, pad1} present investigations on the excitation probability evaluated in a finite time interval for Unruh-DeWitt detectors. The paper is organized as follows. In Sec.~II we discuss the Hamiltonian of two identical two-level atoms weakly coupled with a massless scalar field in Minkowski vacuum. Our ``atoms" can be seen as a pair of standard Unruh-DeWitt detectors. In Sec.~III we evaluate the associated response functions. In Sec.~IV we examine with more detail the different behaviors of the so-called crossed response functions for the transition rate to the symmetric entangled state. We briefly discuss on the total transition rate for equal accelerations of the atoms. In Sec.~V we study the mean life of the symmetric entangled state. In Sec.~VI we present a discussion of the obtained results. Henceforth we work with units such that $\hbar =c=k_{B}=1$ and the Minkowski metric we use is given by $\eta _{\mu \nu}=\mathrm{diag} (-1,1,1,1)$.

\section{Two identical atoms coupled with a massless scalar field}

Let us consider two identical two-level atoms interacting with a massless scalar field in a four-dimensional Minkowski space-time. Here we consider that the atoms are moving along different hyperbolic trajectories. Let us first establish the dependence between the proper times by the structure of Rindler coordinates
\begin{equation}
t=a^{-1}e^{a\xi} \sinh(a\eta), \qquad z=a^{-1}e^{a\xi} \cosh(a\eta),
\label{rindler}
\end{equation}
see, for instance, Fig.~$13$ and discussion from Ref.~\cite{birrel}. Lines of constant $\eta$ are straight lines ($z\sim t$), whereas the lines of constant $\xi$ are hyperbolae $z^{2}-t^{2}=a^{-2}e^{2a\xi}\equiv \textrm{constant}$. As such, they represent the world lines of uniformly accelerated observers in Minkowski space-time. The proper acceleration is defined by $ae^{-a\xi}=\alpha ^{-1}$, and the proper time $\tau$ of the uniformly accelerated observer is related to $\xi$ and $\eta$ by $\tau = e^{a\xi}\eta$. Notice that two different hyperbolae in the same Rindler wedge, represented by different constants $\xi_1, \xi_2$ and parametrized by different proper times $\tau_1, \tau_2$, respectively, can be cut by a single line of constant $\eta$. This means that such proper times can be related as $\tau_1 e^{-a\xi_1} = \tau_2 e^{-a\xi_2}$. We will fully exploit this feature in due course.

In this work we adopt a conventional picture for the uniformly accelerated trajectories of the atoms, which we take to be in the same Rindler wedge $x > |t|$. We are considering that the atoms have different proper accelerations $\alpha _{j}^{-1} = ae^{-a\xi_{j}}$, $j = 1, 2$, and as a consequence they move along world lines $x^{\mu}(\tau_1) = (t(\tau_1),x_1,y_1,z(\tau_1))$ and $x^{\mu}(\tau_2) = (t(\tau_2),x_2,y_2,z(\tau_2))$ parametrized by different proper times. The functions $t(\tau_{j})$ and $z(\tau_{j})$, $j=1,2$, are given by expressions analogous to Eq.~(\ref{rindler}), the difference being that for each atom one has a different $\xi$ and a different $\tau$. The orthogonal distance between the atoms is given by the quantity $|\Delta \mathbf{x}| = \sqrt{(x_{2}-x_{1})^{2}-(y_{2}-y_{1})^{2}}$, assumed to be a constant ({\it orthogonal} here means perpendicular to the $(t,z)$-plane). As a consequence identical accelerations and no orthogonal separation simply imply coinciding world lines. 

As mentioned above, because the atoms are moving along distinct hyperbolic trajectories, one has two proper times. However, to derive the time-evolution operator associated with the coupled system, and then calculate transition probabilities, one must choose a common time variable. This implies that a certain care is mandatory when deciding which parameter will be used to describe the time evolution of the system. In this paper we follow a similar recipe as undertaken in Ref.~\cite{cliche1} and refer to one of the atom's proper time $\tau_1$ and then relate the other proper time $\tau_2$ to $\tau_1$. As asserted above, this can be easily achieved; the dependence between the proper times is given by the lines of constant $\eta$, such that
\begin{equation}
\tau _{2}(\tau _{1})=\tau _{1}e^{a(\xi _{2}-\xi _{1})},
\end{equation}
with $e^{a(\xi _{2}-\xi _{1})}=\alpha _{2}/\alpha _{1}$. We will also analyze the case of equal proper accelerations $\alpha_1 = \alpha_2$ but always keeping a non-vanishing constant orthogonal distance 
$|\Delta \mathbf{x}|$. For an analysis of the interesting situation in which the atoms do not belong to the same Rindler wedge, and with (anti-) parallel trajectories, we refer the reader to Ref.~\cite{Salton:15}.

The total Hamiltonian of the system with respect to the coordinate time $t$ can be written as
\begin{equation}
H = H_{A} + H_{F}+ H_{int},
\end{equation}
where $H_{A}$ is the free atomic Hamiltonian, $H_{F}$ is the free field Hamiltonian and $H_{int}$ describes the interaction between the atoms and the fields. Let us briefly discuss each of such terms. We may express the atomic Hamiltonian in the Dicke notation as~\cite{dick1}
\begin{equation}
H_{A}=\frac{\omega _{0}}{2}\left[ S_{1}^{z} \otimes {\rm 1\!l} _{2}\,\frac{d\tau_1(t)}{dt} + {\rm 1\!l} _{1}\otimes S_{2}^{z}\,\frac{d\tau_2(t)}{dt} \right]
\label{hatom}
\end{equation}
where $S_{i}^{z}= | e _{i}\rangle \langle e _{i}| - |g _{i}\rangle \langle g _{i}|$ is associated with the i-th atom and $\left| g _{i}\right\rangle $, $\left| e _{i}\right\rangle $ is the ground and excited state of the i-th atom, respectively. The eigenstates and respective energies are given by
\begin{equation*}
E_{gg}=-\omega _{0} \quad \left| g g\right\rangle = \left| g _{1}\right\rangle \left| g _{2}\right\rangle ,
\end{equation*}
\begin{equation*}
E_{ge}=0 \quad \left| g e\right\rangle = \left| g _{1}\right\rangle \left| e _{2}\right\rangle ,
\end{equation*}
\begin{equation*}
E_{eg}=0 \quad \left| e g\right\rangle = \left| e _{1}\right\rangle \left| g _{2}\right\rangle ,
\end{equation*}
\begin{equation}
E_{ee}=\omega _{0} \quad \left| e e\right\rangle = \left| e _{1}\right\rangle \left| e _{2}\right\rangle ,
\label{energies1}
\end{equation}
where a tensor product is implicit. Another possible choice is the Bell-state basis. The Bell states are known as the four maximally entangled two-qubit states. In terms of the above product states, the Bell states are expressed as
\begin{equation}
\left| \Psi ^{\pm} \right\rangle =\frac{1}{\sqrt{2}}\left( \left| g _{1}\right\rangle \left| e _{2}\right\rangle \pm \left| e _{1}\right\rangle \left| g _{2}\right\rangle \right) ,
\label{bellstate1}
\end{equation}
\begin{equation}
\left| \Phi ^{\pm} \right\rangle =\frac{1}{\sqrt{2}}\left( \left| g _{1}\right\rangle \left| g _{2}\right\rangle \pm \left| e _{1}\right\rangle \left| e _{2}\right\rangle \right) .
\label{bellstate2}
\end{equation}
The Hamiltonian (\ref{hatom}) is showing a degeneracy associated with the eigenstates $\left| g e\right\rangle$ and $\left| e g\right\rangle$. Any linear combination of these degenerate eigenstates will be an eigenstate of the atomic Hamiltonian and these linear combination must have the same degenerate energy value. Then (\ref{bellstate1}) are eigenstates of $H_{A}$. On the other hand, it is clear that states given by Eq.~(\ref{bellstate2}) do not represent eigenstates of the atomic Hamiltonian. Such states will not be considered in our analysis, and we are only interested in the investigations concerning the transitions to (or from) the entangled states $\left| \Psi^{\pm} \right\rangle$.

The free Hamiltonian of the quantum field which governs the field time evolution is given by
\begin{equation}
H_{F}=\frac{1}{2}\int d^{3}x \left[ (\dot{\varphi}(x))^{2}+\left( \nabla \varphi (x)\right) ^{2}\right],
\end{equation}
where the dot represents the derivative with respect to $t$. 

At this point we remark that we will be working in the interaction representation. Hence we assume that the coupling between the atoms and the field is described by a monopole interaction in the form 
\begin{equation}
H_{int}(t) = \sum _{j=1}^{2} \mu _{j} \chi_{j}(\tau_{j}(t)) m^{(j)}(\tau_{j}(t)) \varphi [x_{j}(\tau _{j}(t))]\frac{d \tau _{j}(t)}{dt},
\label{hint}
\end{equation}
where 
\begin{equation}
m^{(j)}(\tau _{j}(t)) = |e^{(j)}\rangle \langle g^{(j)}| \, e^{i\omega_{0}\tau_{j}(t)} 
+ | g^{(j)}\rangle \langle e^{(j)}| \, e^{-i\omega_{0}\tau_{j}(t)}
\end{equation}
is the monopole moment operator of the $j$-th atom. For investigations discussing physical motivations of the Unruh-DeWitt detector model as a model for light-matter interactions, we refer the reader to Refs.~\cite{Eduardo:13,Eduardo:14,Eduardo:15,Eduardo:16} and references cited therein. The quantity $\mu _{j}$ in Eq.~(\ref{hint}) is the dimensionless coupling constant of the $j$-th atom and $\varphi [x_{j}(\tau _{j}(t))]$ is the field at the point of the $j$-th atom. The real-valued switching functions $\chi_{j}$ stipulate how the interaction between atoms and fields is turned on and off. The basic assumption one usually assumes is that $\chi_{j}$ have compact support or else have suitably strong fall-off properties in such a way that the system can be envisaged as asymptotically uncoupled in the distant past and future. Henceforth we set $\mu_1 = \mu_2 = \mu$ and we assume that $\mu$ is small.

As mentioned above, we consider that the atoms are moving along world lines $x^{\nu}_{1,2}(\tau _{1,2})$ which are parametrized by the proper times $\tau _{1,2}$. Since in this case the proper times of the atoms do not coincide, we write the time-evolution operator as~\cite{cliche1}
\begin{eqnarray}
U &=&\, T\exp\left\{ -i \int_{-\infty}^{\infty} d\tau _{1} \, \mu \, \Biggl[\chi_{1}(\tau _{1})\,m^{(1)}(\tau _{1})\varphi \left( x^{\nu}_{1}(\tau _{1})\right) \right.
\nonumber \\
&+&\, \left. \chi_{2}(\tau _{2}(\tau _{1}))\, m^{(2)}(\tau _{2}(\tau _{1}))\varphi \left( x^{\nu}_{2}(\tau _{2}(\tau _{1}))\right) \frac{d\tau _{2}(\tau _{1})}{d\tau _{1}}\Biggr] \right\}.
\label{uni}
\end{eqnarray}
where $T$ is the usual time-ordering symbol. Let us consider that the system is initially prepared in the state $\left|gg; 0_{M}\right\rangle$, where $\left| gg\right\rangle$ is the collective ground state for the atoms and $\left| 0_{M}\right\rangle$ is the Minkowski vacuum state of the scalar field. The amplitude in first-order perturbation theory for a transition to the final state 
$|\omega ;\Omega\rangle$ is given by
\begin{eqnarray}
\hspace{-5mm}
A_{\left| gg; 0_{M}\right\rangle \rightarrow  \left| \omega ;\Omega\right\rangle} &=& 
\langle \omega ;\Omega| U | gg;0_{M}\rangle
= - i\mu \bigg\langle \omega ;\Omega\bigg|
\int_{-\infty}^{\infty} d\tau _{1} \left[\chi_{1}(\tau _{1})\,m^{(1)}(\tau _{1})\varphi \left( x^{\nu}_{1}(\tau _{1})\right) \right.
\nonumber\\
&+& \left. \chi_{2}(\tau _{2}(\tau _{1}))\,m^{(2)}(\tau _{2}(\tau _{1}))\varphi \left( x^{\nu}_{2}(\tau _{2}(\tau _{1}))\right) \frac{d\tau _{2}(\tau _{1})}{d\tau _{1}}\right] \bigg| gg;0_{M}\bigg\rangle.
\end{eqnarray}
We are interested in the situation in which $|\omega\rangle$ is taken to be an energy eigenstate of $H_{A}$. In any case, the transition probability to all possible $|\omega ;\Omega\rangle$ up to order $\mu^2$ is obtained by squaring the modulus of the above expression and summing over $\omega$ and the complete set $\Omega$. In general, the sum over a complete set of field states in this case will produce terms such as
\bea
&& \sum_{\Omega} \langle 0_{M}| \phi(x_1) \phi(x_2) \ldots \phi(x_n) |\Omega\rangle 
\langle \Omega| \phi(y_1) \phi(y_2) \ldots \phi(y_m) |0_{M}\rangle
\nn\\
&& ~~~~~~~~~~~~~~~~~~~~~~~~~~~~~~~~~~~~~~~~~~~~~~~~~~~~~~ 
= \langle 0_{M}| \phi(x_1) \phi(x_2) \ldots \phi(x_n) \phi(y_1) \phi(y_2) \ldots \phi(y_m) |0_{M}\rangle,
\eea
in other words, tracing out of field degrees of freedom yields vacuum expectation values of products of quantum fields. In the above equation $x_1, x_2, \ldots$ represent generic space-time points (obviously $n, m$ are positive integers). In our case one should properly take into account the dependence on proper time for each space-time point in such an expression (e.g., $x_1 = x_1(\tau_1)$, $x_2 = x_2(\tau_2)$, etc.) since the coupling between atoms and the quantum field is effective only on the trajectories of the atoms. 

For a general final atomic state $|\omega\rangle$ (including a generic entangled state), one should also consistently take into account the ${\cal O}(\mu^2)$ terms in the above amplitude so that one obtains a correct probability transition up to order $\mu^2$. However, since we are solely interested in the probability associated with the particular transition $\left| \omega ' \right\rangle \rightarrow \left| \omega \right\rangle$, where $|\omega '\rangle = | gg\rangle$ and $\left| \omega \right\rangle$ is one of the maximally entangled states $\left| \Psi ^{\pm} \right\rangle$, contributions of order $\mu^2$ in the transition probability will be produced only by the ${\cal O}(\mu)$ terms in the amplitude. In first-order approximation the transition probability is then given by
\begin{equation}
\Gamma _{\left| \omega ' \right\rangle \rightarrow \left| \omega \right\rangle} (\Delta \omega, \Delta t) 
= \mu^{2}\sum _{i,j} \left[m^{(i)*}_{\omega \omega '}m^{(j)}_{\omega \omega '}
F_{ij}(\Delta \omega, \Delta t)\right],
\label{sumaf}
\end{equation}
$i=1,2$, $j=1,2$, where $\Delta \omega = \omega - \omega '$ is the energy gap between the eigenstates $|\omega\rangle$ and $|\omega' \rangle$, respectively, and the matrix elements are given by
\begin{equation*}
m_{\omega \omega '}^{(1)}=\left\langle \omega \right| m ^{(1)}(0) \otimes {\rm 1\!l} _{2} \left| \omega ' \right\rangle
\end{equation*}
\begin{equation*}
m_{\omega \omega '}^{(2)}=\left\langle \omega \right| {\rm 1\!l} _{1}  \otimes m ^{(2)}(0) \left| \omega ' \right\rangle .
\end{equation*}
The quantity $\Delta t$ will be defined in detail below but for now we can note that it is related to the characteristic time scale of the switching function 
\footnote{The parameter $\Delta t$ has a different specification depending on the switching function considered. As will be subsequently discussed, in this paper we choose two different descriptions for the switching function, namely, a sharp switching function and a Gaussian switching function. The usage of such denominations will be clarified in the next section. In any case, for the former, one has that $\Delta t = \tau_f - \tau_0$, where $\tau_0$ and $\tau_f$ are related to a finite interval of interaction between the atom and the quantum field. On the other hand, for the Gaussian switching function, one simply has that $\Delta t = \sqrt{2\pi} \zeta$, where the physical interpretation for the parameter $\zeta$ is discussed later on in this work. Hence, one notices that any detailed specification for the parameter $\Delta t$ accordingly requires an elaborated identification of the particular switching function.}. 
As discussed above $|\omega '\rangle = | gg\rangle$ and $\left| \omega \right\rangle = \left| \Psi ^{\pm} \right\rangle$, which gives $\Delta \omega = \omega_0$, the energy gap between the states $\left| \Psi ^{\pm} \right\rangle$ and $| gg\rangle$. The corresponding response functions are defined by
\begin{eqnarray}
F_{ij}(\Delta \omega, \Delta t) &=& \int_{-\infty}^{\infty} d\tau _{1} 
\int_{-\infty}^{\infty} d\tau _{1}'e^{-i\Delta \omega (\tau _{i}(\tau _{1})-\tau _{j}(\tau _{1}'))}\,\chi_{i}(\tau _{i}(\tau _{1}))\,\chi_{j}(\tau _{j}(\tau _{1}'))
\nonumber\\
&\times&\,G_{ij}^{+}(\tau_1,\tau_1') \frac{d\tau _{j}(\tau' _{1})}{d\tau' _{1}}\frac{d\tau _{i}(\tau _{1})}{d\tau _{1}}.
\end{eqnarray}
Here $G_{ij}^{+}(\tau_1,\tau_1')=\left\langle 0_{M}\right| \varphi(x_{i}(\tau_{i}(\tau_{1})))
\varphi(x_{j}(\tau_j(\tau _{1}'))) \left| 0_{M}\right\rangle $ is the positive-frequency Wightman function in Minkowski space-time for a massless scalar field, which is given by
\begin{equation}
G_{ij}^{+}(\tau,\tau') = \frac{1}{8\pi ^{2}}\frac{1}{X(\tau,\tau')},
\label{wigh}
\end{equation}
where $X (\tau ,\tau ')$ is given by
\begin{eqnarray}
2 X(\tau,\tau') &=& (x_{i}(\tau) - x_{j}(\tau'))^2
\nonumber\\
&=&\, -(t_{i}(\tau)-t_{j}(\tau')-i\epsilon )^2
+|\mathbf{x}_{i}(\tau)-\mathbf{x}_{j}(\tau')|^{2}.
\end{eqnarray}
Note that we are employing the standard $i\epsilon$ prescription for the Wightman function. 

We are interested in the transition probability to an entangled state for a pair of atoms travelling in different hyperbolic world lines. Hence we study the transition $\left| gg \right\rangle \rightarrow \left| \Psi ^{\pm} \right\rangle$, with $\Delta \omega = \omega _{0}> 0$. The appearance of cross terms in the transition probability has its origin in the fact of working with two atoms, both interacting with a common scalar quantum field.

We may define the associated transition rate as follows
\begin{equation}
\mathcal{R} _{\left| \omega ' \right\rangle \rightarrow \left| \omega \right\rangle}(\Delta \omega ,\Delta t) =\frac{\partial\Gamma _{\left| \omega ' \right\rangle \rightarrow \left| \omega \right\rangle} (\Delta \omega ,\Delta t)}{\partial(\Delta t)},
\label{rtotal1}
\end{equation}
where $\Gamma _{\left| \omega ' \right\rangle \rightarrow \left| \omega \right\rangle} (\Delta \omega ,\Delta t)$ is given by Eq.~(\ref{sumaf}). 

The usage of the terminology ``transition rate'' throughout the paper calls for some clarification.  We here employ the standard parlance in order to make contact with usual expositions encountered in the context of quantum mechanical time-dependent perturbation theory. Accordingly, the word ``transition" in the present scenario is connected with the possibility of finding the system, in a later time, in a state which is different from the initial state it was prepared in due to the perturbation introduced by the interaction of the atoms with the quantum field, as a result of its unitary evolution described by a time-evolution operator. Hence the utilization of the expression ``transition rate'' is justified in the sense just described. Nevertheless, one must be truly cautious when referring to the quantity defined in Eq.~(\ref{rtotal1}) as a ``transition rate'' when considering the transition $\left| gg \right\rangle \rightarrow \left| \Psi ^{\pm} \right\rangle$. Indeed, a probability is ultimately used in quantum mechanics in the context of a measurement. In the specific case of the transition $\left| gg \right\rangle \rightarrow \left| \Psi ^{\pm} \right\rangle$, we remind the reader that the detection of the entangled final state requires a non-local projective measurement on both atoms.

In order to present an explicit expression for the transition rate as defined above, one needs to properly evaluate in detail each of the response functions $F_{ij}(\Delta \omega, \Delta t)$. This is the topic of the next section.

\section{Discussion on the response functions}

\subsection{Individual response functions, $F_{11}(\Delta \omega ,\Delta t)$ and $F_{22}(\Delta \omega ,\Delta t)$}

As a first analysis, we will consider switching functions with a sharp switch-on and switch-off as given below:
\begin{equation}
\chi_{1}(\tau_{1}) = \Theta(\tau_{1} - \tau_0)\,\Theta(\tau_{f} - \tau_{1})
\label{sharp1}
\end{equation}
and
\begin{equation}
\chi_{2}(\tau_{2}(\tau_{1})) = \Theta\left[\frac{\alpha_1}{\alpha_2}\,\tau_{2}(\tau_{1}) - \tau_0\right]\,
\Theta\left[\tau_{f} - \frac{\alpha_1}{\alpha_2}\,\tau_{2}(\tau_{1})\right]
\label{sharp2}
\end{equation}
where $\Theta(x)$ is the usual Heaviside step function and $\tau_0$ and $\tau_f$ are real parameters associated with a finite interval of interaction between the atom and the quantum field (we always take $\tau_f > \tau_0$). In this case the response function corresponds to sharply switching on the interaction at $\tau_0$ and sharply switching it off at $\tau_f$. Let us first evaluate the contribution $F_{11}(\Delta \omega)$ to the total response function. Inserting Eq.~(\ref{rindler}) in expression~(\ref{wigh}) one finds all the relevant positive-frequency Wightman functions of the problem. In particular, the Wightman function $G_{11}^{+}$ is given by
\begin{equation}
G_{11}^{+}\left(\tau _{1} - \tau _{1}' \right) =-\frac{1}{16\pi ^{2}\alpha ^{2}_{1}
\sinh ^{2}\left(\frac{\tau _{1}-\tau '_{1}}{2\alpha _{1}}-\frac{i\epsilon}{\alpha _{1}}\right)} .
\label{g11i}
\end{equation}
In order to reach the above result one should change the interpretation of the regulator $\epsilon$. As discussed in Ref.~\cite{hu1}, the $\epsilon$ in Eq.~(\ref{g11i}) is to be regarded as a physical cutoff corresponding to the time resolution of the detectors and in principle it should not be taken the zero limit when the observational time interval of the system under study is not large.

Using known series identities~\cite{series1} we can rewrite (\ref{g11i}) as
\begin{equation}
G_{11}^{+}(\tau _{1}-\tau_{1}') =-\frac{1}{4\pi ^{2}} \sum _{n=-\infty}^{\infty} \left( (\tau _{1}-\tau _{1}')-2i\epsilon +2\pi i \alpha _{1}n \right) ^{-2}.
\label{g11ii}
\end{equation}
In this way, changing variables to
\begin{equation}
\psi = \tau _{1}-\tau _{1}' \qquad \eta = \tau _{1}+\tau _{1}',
\label{change}
\end{equation}
we have,
\begin{equation}
F_{11}(\Delta \omega ,\Delta t) = \int _{-\Delta t}^{\Delta t}d\psi \left(\Delta t - |\psi| \right) e^{-i\Delta \omega \psi} G_{11}^{+}(\psi),
\label{f11integral}
\end{equation}
where $\Delta t = \tau_{f} - \tau_0$. The evaluation of the integral leads us to~\cite{nami1}
\begin{eqnarray}
 F_{11}(\Delta \omega ,\Delta t) &=& \frac{\Delta t}{2\pi ^{2}}
\left\lbrace e^{-2\epsilon|\Delta \omega|}\pi |\Delta \omega |\Theta (-\Delta \omega)\left[1 + \frac{1}
{e^{2\pi \alpha _{1}|\Delta \omega |}-1}\right] \right.
\nonumber\\
&+&\, \left. \frac{e^{2\epsilon\Delta \omega}\pi\Delta \omega\Theta (\Delta \omega)}{e^{2\pi \alpha _{1}\Delta \omega }-1}
+ {\cal P}_{1}(\Delta\omega, \Delta t, \epsilon) + \Re\left[{\cal J}_{1}(\Delta\omega, \Delta t, \epsilon)\right]\right\rbrace
\nonumber \\
&+&\,  \frac{1}{2\pi ^{2}}\biggl\{{\cal P}_{2}(\Delta\omega, \Delta t, \epsilon) 
+ \Re\left[{\cal J}_{2}(\Delta\omega, \Delta t, \epsilon)\right]\biggr\}.
\label{f11tiempofinito}
\end{eqnarray}
For details concerning such a calculation and the associated definitions of the functions ${\cal P}_{k}, {\cal J}_{k}$, $k = 1, 2$, we refer the reader the~\ref{app1}. We simply observe that the function ${\cal P}_2$ generates the familiar logarithmic divergence in the limit $\epsilon \to 0$~\cite{nami1}. We are interested in the rate
\begin{equation}
R_{ij}(\Delta \omega ,\Delta t)=\frac{\partial F_{ij}(\Delta \omega ,\Delta t)}{\partial(\Delta t)}
\label{rate1}
\end{equation}
which is related to the mean life of states. From the expression~(\ref{r11sec1}) found in~\ref{app1} it is easy to see that, for large observational time intervals as compared with the time resolution of the detectors, one may safely take the limits 
$\Delta t \rightarrow \infty$ and $\epsilon \to 0$. We obtain the following expression
\begin{eqnarray}
\lim_{\epsilon \to 0} \lim _{\Delta t \rightarrow \infty} R_{11}(\Delta \omega ,\Delta t) &=& \frac{|\Delta \omega |}{2\pi}\biggl\lbrace \Theta (-\Delta \omega) \left[1+\frac{1}{e^{2\pi \alpha _{1}|\Delta \omega|}-1}\right]
+ \Theta (\Delta \omega)\frac{1}{e^{2\pi \alpha _{1}\Delta \omega}-1}\biggr\rbrace.
\label{r11l}
\end{eqnarray}
The above equation shows us that the equilibrium between the uniformly accelerated atom and scalar field in the Minkowski vacuum state $\left| 0 _{M}\right\rangle$ is the same as that which would have been achieved had this atom followed an inertial trajectory but immersed in a bath of thermal radiation at the temperature $\beta ^{-1}_{1}=1/(2\pi \alpha _{1})$.

Let us now consider the response function $F_{22}$. The associated Wightman function is given by
\begin{equation}
G_{22}^{+}\left(\tau _{2}(\tau _{1}),\tau _{2}(\tau _{1}')\right) =-\frac{1}{16\pi ^{2}\alpha ^{2}_{2}\sinh ^{2}\left(\frac{\tau _{1}-\tau '_{1}}{2\alpha _{1}}-\frac{i\epsilon}{\alpha _{1}}\right)}.
\label{f22i}
\end{equation}
As above we have considered a modification in the interpretation of the regulator $\epsilon$. Performing similar steps as before, we can rewrite (\ref{f22i}) as
\begin{equation}
G_{22}^{+}(\psi) =-\frac{1}{4\pi ^{2}} \frac{\alpha _{1}^{2}}{\alpha _{2}^{2}} \sum _{n=-\infty}^{\infty} \left( \psi -2i\epsilon +2\pi i \alpha _{1}n \right) ^{-2},
\end{equation}
such that, the expression for the response function $F_{22}(\Delta \omega ,\Delta t)$ yields
\begin{eqnarray}
F_{22}(\Delta \omega ,\Delta t) &=&\, e^{2a(\xi _{2}-\xi _{1})}\int _{-\Delta t}^{\Delta t}d\psi\,
(\Delta t -|\psi |) e^{-i\Delta \omega e^{a(\xi _{2}-\xi _{1})} \psi}G_{22}^{+}(\psi).
\label{f22int}
\end{eqnarray}
Comparing Eqs.~(\ref{f22int}) and~(\ref{f11integral}), it is straightforward to see that the associated expressions for $F_{22}(\Delta \omega ,\Delta t)$ and also $R_{22}(\Delta \omega ,\Delta t)$ can be easily derived from the analogous results for the atom $1$ by performing the replacement $\Delta\omega \to (\alpha_{2}/\alpha_{1})\Delta\omega$ (this amounts to the inclusion of the usual redshift factor). We observe that, for large observational time intervals (as compared with $\epsilon$), the equilibrium between the atom $2$ and scalar field in the Minkowski vacuum state is the same as the equilibrium of this atom in an inertial trajectory and a bath of thermal radiation at the temperature $\beta ^{-1}_{2} = 1/(2\pi \alpha _{2})$.

A brief digression on the use of the above switching functions is in order. In the first place, had we not considered an alteration in the character of the regulator $\epsilon$, we would not be able to derive expressions~(\ref{g11i}) and~(\ref{f22i}) for the Wightman functions. For instance, the function $G_{11}^{+}$ would look like
\beq
G_{11}^{+}\left(\tau _{1}, \tau _{1}' \right) = \frac{1}{4\pi ^{2}}
\frac{1}{- 4 \alpha^{2}_{1}\sinh\left(\frac{\tau _{1} - \tau _{1}'}{2\alpha_{1}}\right)
\left[\sinh\left(\frac{\tau _{1} - \tau _{1}'}{2\alpha_{1}}\right) 
-  i (\epsilon'/\alpha_{1})\cosh\left(\frac{\tau _{1} + \tau _{1}'}{2\alpha_{1}}\right)\right]+\epsilon'^2 }.
\label{math}
\eeq
A similar expression holds for $G_{22}^{+}$. The $\epsilon'$ in the above equation does not have the same status as the $\epsilon$ in Eqs.~(\ref{g11i}) and~(\ref{f22i}); here it is a mathematical cut-off and the limit $\epsilon' \to 0$ could be taken at the end of the calculations. However, this has far-reaching consequences. As discussed by~\cite{Schlicht}, with such a mathematical definition for the regulator one obtains unphysical results in the situation where the detectors are undergoing uniform acceleration (e.g., non-Lorentz invariant terms in the transition rate). On the other hand, as pointed out by Satz, the origin of such manifestly physically unreasonable outcomes can be traced back to the assumption of a sudden switch-on and switch-off of the detectors~\cite{satz}. By conceiving the following smooth switching function 
$$
\chi(u) = {\cal Q}_{1}\left(\frac{u - \tau_0 - \delta}{\delta}\right)\,{\cal Q}_{2}\left(\frac{-u + \tau + \delta}{\delta}\right),
$$
with $\tau > \tau_0$ and $\delta > 0$, he concludes that the results for the transition rate obtained from the standard regularization are both finite and Lorentz invariant. The functions ${\cal Q}_{1,2}$ are smooth switching functions obeying 
${\cal Q}_{1,2}(x) = 0$ for $x < 0$ and ${\cal Q}_{1,2}(x) = 1$ for $x > 1$. In this case $\delta$ plays the same role as the regulator $\epsilon$ in Eqs.~(\ref{g11i}) and~(\ref{f22i}). Indeed, Satz finds an identical logarithmic divergence as in Eq.~(\ref{f11tiempofinito}) with $\epsilon \to 0$ (the case for $G_{22}^{+}$ leads to the same divergence). As remarked in the aforementioned reference, the transition rate that follows in the limit of sharp switching is well defined, yet it holds only as a valid approximation for fixed energy gaps when the observational time interval $\tau - \tau_0$ is much longer than $\delta$. Clearly a similar situation emerges in our analysis. Hence based on such arguments, we infer that our results remain valid only when $\Delta t \gg \epsilon$, with the given energy gap not too large. The physical reasonableness of our model is thus warrant and such considerations corroborate our findings. Incidentally, this also shows why $\epsilon$ (as well as $\Delta t$) must be kept finite; this ensures that the physical input on the detectors do not conflict with the usual conditions of the Wightman distribution function.

In order to bring forth a different picture for the contention just exposed, let us proceed as follows. We now assume smooth switching functions in the form of Gaussian profiles~\cite{emm}:
\begin{equation}
\chi_{1}(\tau_{1}) = e^{-(\tau_{1} - \tau)^2/\zeta^2}
\label{gauss1}
\end{equation}
and
\begin{equation}
\chi_{2}(\tau_{2}(\tau_{1})) = e^{-\left[\frac{\alpha_1}{\alpha_2}\,\tau_{2}(\tau_{1}) - \tau\right]^2/\zeta^2}.
\label{gauss2}
\end{equation}
Now the response function is associated with a smooth switching on and off of the coupling between the atoms and the quantum field. The parameters $\tau$ and $\zeta$  should be judiciously chosen in order to provide a smooth approximation to the previous sharp switching functions; $\sqrt{2\pi}\,\zeta$ approximately mimics the preceding $\Delta t$ and is interpreted as the characteristic timescale of the switching function. In turn, $\tau$ embodies the analogous corresponding character displayed by the quantity $T = (\tau_{f} + \tau_{0})/2$ in the sharp switching function. The Gaussian form for the switching function will prove to be useful in order to provide another physical support for our results. As we shall see, this will make clear the distinctions between the properties of the response function which depend on the trajectory of the detectors (as well as the quantum state of the field), and those features which are result of the sharp switching assumption.

By taking into account Eqs.~(\ref{gauss1}) and~(\ref{gauss2}) one obtains
\begin{equation}
F_{11}(\Delta \omega ,\zeta) = -\frac{\sqrt{2 \pi } \zeta}{32\pi ^{2}\alpha_1^2}
\int _{-\infty}^{\infty}d\psi \frac{e^{-i\Delta \omega \psi}\, e^{-\frac{\psi^2}{2 \zeta^2}}}
{\sinh ^{2}\left(\frac{\psi - 2 i \epsilon}{2\alpha _{1}}\right)}.
\label{gauss3}
\end{equation}
As before the expression for $F_{22}(\Delta \omega ,\Delta t)$ can be derived from the above equation by performing the replacement $\Delta\omega \to (\alpha_{2}/\alpha_{1})\Delta\omega$. Observe that we are using $\epsilon$ as a physical cutoff as explained above, i.e., the Wightman function $G_{11}^{+}$ considered is given by Eq.~(\ref{g11i}). We are particularly interested in two asymptotic limits for $\zeta$. First, let us consider that $\zeta \lesssim \epsilon \ll 1$. In this limit one may approximately take
$$
e^{-\frac{\psi^2}{2 \zeta^2}} \approx \sqrt{2\pi}\,\zeta\,\delta(\psi).
$$
Hence
\bea
F_{11}(\Delta \omega ,\zeta) &\approx& \frac{\zeta^2}{16\pi\alpha_1^2}
\frac{1}{\sin^{2}(\epsilon/\alpha _{1})}
\nn\\
&\approx&\,\frac{1}{16\pi}\,\frac{\zeta^2}{\epsilon^2} + \cdots
\eea
where in the last line we have considered a standard Taylor series expansion for the sine function. Note that the first term in the above series expansion coincides (up to an irrelevant multiplicative constant) with the result obtained from the Wightman function written in the form~(\ref{math}) by taking $\epsilon' = \epsilon$. So the behavior of the individual response functions for small $\zeta$ is roughly the same for both definitions of the regulator $\epsilon$; incidentally both are finite for any small values of $\epsilon$ within the condition $\zeta \lesssim \epsilon$.

The other interesting limit is when $\zeta \gg \epsilon$. In order to perform a detailed analysis for this situation we employ a Taylor series expansion for the Gaussian exponential in Eq.~(\ref{gauss3}). By resorting to the Lebesgue's dominated convergence theorem, one gets
\bea
\hspace{-4mm}
F_{11}(\Delta \omega ,\zeta) &=& -\frac{\sqrt{2 \pi } \zeta}{32\pi ^{2}\alpha_1^2}
\sum_{k=0}^{\infty}\,\frac{(-1)^{k}}{(2 \zeta^2)^{k}\,k!}
\int _{-\infty}^{\infty}d\psi \frac{e^{-i\Delta \omega \psi}\, \psi^{2k}}
{\sinh ^{2}\left(\frac{\psi - 2 i \epsilon}{2\alpha _{1}}\right)}
\nn\\
&=&\,\frac{\sqrt{2 \pi } \zeta}{4\pi}
\left\{\Theta(\Delta\omega)\,\exp{\left(\frac{{\cal D}(\Delta\omega)}{2\zeta^2}\right)}
\left[\Delta\omega  e^{2 \Delta\omega  \epsilon}\,\frac{1}{e^{2 \pi  \alpha _1 \Delta\omega }-1}\right]
\right.
\nn\\
&+& \left. \Theta(-\Delta\omega)\,\exp{\left(\frac{{\cal D}(|\Delta\omega|)}{2\zeta^2}\right)}
\left[|\Delta\omega|  e^{- 2 |\Delta\omega| \epsilon}
\left(1 + \frac{1}{e^{2 \pi  \alpha _1 |\Delta\omega| }-1}\right)\right]\right\}
\label{gauss-f11}
\eea
where ${\cal D}(y) = \partial^{2}/\partial y^2$ and in the last line we have employed the residue theorem in order to solve the integral. In the asymptotic limit $\zeta \gg \epsilon$ one may consider only the leading term in the above expansion. Hence, if one agrees with the identification $\Delta t = \sqrt{2 \pi } \zeta$ one finds, for the associated transition rate
\begin{eqnarray}
\lim_{\epsilon \to 0} \lim _{\Delta t \rightarrow \infty} R_{11}(\Delta \omega ,\Delta t) &\approx& \frac{|\Delta \omega |}{4\pi}\biggl\lbrace \Theta (-\Delta \omega) \left[1+\frac{1}{e^{2\pi \alpha _{1}|\Delta \omega|}-1}\right]
+ \Theta (\Delta \omega)\frac{1}{e^{2\pi \alpha _{1}\Delta \omega}-1}\biggr\rbrace.
\end{eqnarray}
We notice that such an expression, apart from an unimportant factor of $2$ in the denominator, is identical to Eq.~(\ref{r11l}).

The conclusions one can draw from the above considerations are obvious. For small values of $\zeta$, both Wightman functions~(\ref{g11i}) and~(\ref{math}) generate the same individual response functions in the leading order with the Gaussian switching function. In addition, both are finite for a finite $\epsilon$. On the other hand, in the asymptotic limit $\zeta \gg \epsilon$ the Gaussian and the sharp switching functions produce the same behavior for the transition rates. However, notice that Eq.~(\ref{gauss-f11}) is finite for $\epsilon \to 0$, whereas Eq.~(\ref{f11tiempofinito}) displays a logarithmic divergence in the same limit. Hence, this confirms the previous assertion that the analysis within a sharp switch-on and switch-off of the interaction between detectors and quantum fields are valid as long as $\Delta t \gg \epsilon$, with $\epsilon$ kept finite.

\subsection{Crossed response functions, $F_{12}(\Delta \omega ,\Delta t)$ and $F_{21}(\Delta \omega ,\Delta t)$}

The first two response functions discussed above correspond to individual atomic transitions. Therefore one expects that the response functions $F_{12}(\Delta \omega ,\Delta t)$ and $F_{21}(\Delta \omega ,\Delta t)$ exhibit the existence of cross-correlations between the atoms mediated by the field. In order to unveil such a behavior, we now proceed to evaluate such contributions. 

We begin with the sharp-switching case, Eqs.~(\ref{sharp1}) and~(\ref{sharp2}). It is easy to show that the positive frequency Wightman functions for both cases are equal and are given by
\begin{equation}
G_{21}^{+}\left(\tau _{2}(\tau _{1}),\tau _{1}'\right) = G_{12}^{+}\left(\tau_{1},\tau _{2}(\tau _{1}') \right)
= G^{+}_{c}(\tau _{1}-\tau _{1}')
\end{equation}
where
\begin{equation}
G^{+}_{c}(\tau _{1}-\tau _{1}')= -\frac{1}{16\pi ^{2}\alpha _{1}\alpha _{2}}G^{+}_{c_{0}}(\tau _{1}-\tau _{1}',\epsilon)
\end{equation}
and
\beq
G^{+}_{c_{0}}(\tau _{1} -\tau _{1}',\epsilon) = \left[\sinh\Biggl(\frac{\tau _{1}-\tau _{1}'}{2\alpha _{1}}-\frac{4i\epsilon}{(\alpha _{1}+\alpha _{2})}+\frac{\phi}{2} \Biggr) 
\sinh\left(\frac{\tau _{1}-\tau _{1}'}{2\alpha _{1}}-\frac{4i\epsilon}{(\alpha _{1}+\alpha _{2})}-\frac{\phi}{2} \right)\right]^{-1}
\eeq
with
\begin{equation}
\cosh \phi = 1+\frac{(\alpha _{1}-\alpha _{2})^{2}+|\Delta \mathbf{x}|^{2}}{2\alpha _{1}\alpha _{2}}
\label{cosh}
\end{equation}
with $|\Delta \mathbf{x}|^{2} = (x_{2}-x_{1})^{2}-(y_{2}-y_{1})^{2}$ as given above. As before we have suitably redefined the meaning of the regulator $\epsilon$. One has that:
\beq
F_{21}(\Delta \omega ,\Delta t) = \frac{1}{2}e^{a(\xi _{2}-\xi _{1})}\int _{-\Delta t} ^{\Delta t}d\psi 
\int _{|\psi |+2\tau _{0}} ^{-|\psi |+2\tau _{f}}d\eta\,
e^{-i\Delta \omega (a_{-}/2\alpha _{1})\eta}e^{-i\Delta \omega (a_{+}/2\alpha _{1})\psi}\,G^{+}_{c}(\psi),
\label{expresion11}
\eeq
where we used Eq.~ (\ref{change}) and where $a_{-}=\alpha _{2}-\alpha _{1}$ and $a_{+}=\alpha _{2}+\alpha _{1}$. After some algebraic manipulations, one gets
\begin{eqnarray}
F_{21}(\Delta \omega ,\Delta t) &=& \frac{-i}{16\pi ^{2}\alpha_{1}\Delta \omega a_{-}}
\left\lbrace e^{-i\Delta \omega (a_{-}/2\alpha _{1})(\Delta t + 2 T)} \right. 
\left[I_{\epsilon}(\Delta \omega ,\Delta t,1) + I_{-\epsilon}(\Delta \omega ,\Delta t,-\alpha _{2}/\alpha _{1})\right]
\nonumber \\
 &-&\, \left. e^{i\Delta \omega (a_{-}/2\alpha _{1})(\Delta t- 2 T)}\left[ I_{-\epsilon}(\Delta \omega ,\Delta t,-1)\right.  \left. +I_{\epsilon}(\Delta \omega ,\Delta t,\alpha _{2}/\alpha _{1})\right] \right\rbrace
\label{f21aaa}
\end{eqnarray}
where
\begin{equation}
I_{\epsilon}(\Delta \omega ,\Delta t, \sigma)\equiv \int _{0}^{\Delta t}d\psi e^{-i\sigma \Delta \omega \psi}
G^{+}_{c_{0}}(\psi,\epsilon).
\label{integral1}
\end{equation}
In a similar fashion, one has that
\begin{eqnarray}
F_{12}(\Delta \omega ,\Delta t) &=& \frac{i}{16\pi ^{2}\alpha_{1}\Delta \omega a_{-}}
\left\lbrace e^{i\Delta \omega (a_{-}/2\alpha _{1})(\Delta t + 2 T)} 
\left[I_{-\epsilon}(\Delta \omega ,\Delta t,-1) + I_{\epsilon}(\Delta \omega ,\Delta t,\alpha _{2}/\alpha _{1})\right] \right.
\nonumber \\
&-&\, \left. e^{-i\Delta \omega (a_{-}/2\alpha _{1})(\Delta t- 2 T)}\left[I_{\epsilon}(\Delta \omega ,\Delta t,1) 
+ I_{-\epsilon}(\Delta \omega ,\Delta t,-\alpha _{2}/\alpha _{1})\right] \right\rbrace .
\end{eqnarray}
Observe that $F_{21}(\Delta \omega ,\Delta t)=F_{12}^{*}(\Delta \omega ,\Delta t)$. Hence the object of interest is 
$2\Re[F_{21}(\Delta \omega ,\Delta t)]$. From the results derived in the~\ref{app2}, this is given by
\begin{eqnarray}
\Re[F_{21}(\Delta \omega ,\Delta t)] &=& \frac{1}{4\pi ^{2}\sinh(\phi)\Delta \omega a_{-}}
\nonumber\\
&\times&\,\Biggl\{\cos\left[\frac{\Delta \omega a_{-}}{2\alpha _{1}}(\Delta t + 2 T)\right]
\Re[{\cal A}(\Delta\omega, \alpha_1, \alpha_2) + {\cal I}(\Delta\omega, \Delta t)]
\nn\\
&+&\, \sin\left[\frac{\Delta \omega a_{-}}{2\alpha _{1}}(\Delta t + 2 T)\right]
\Im[{\cal A}(\Delta\omega, \alpha_1, \alpha_2) + {\cal I}(\Delta\omega, \Delta t)]
\nn\\
&+&\, \cos\left[\frac{\Delta \omega a_{-}}{2\alpha _{1}}(\Delta t- 2 T)\right]
\Re[-{\cal A}(\Delta\omega, \alpha_2, \alpha_1) + {\cal I}^{*}(\Delta\omega, \Delta t)]
\nn\\
&-&\, \sin\left[\frac{\Delta \omega a_{-}}{2\alpha _{1}}(\Delta t- 2 T)\right]
\Im[-{\cal A}(\Delta\omega, \alpha_2, \alpha_1) + {\cal I}^{*}(\Delta\omega, \Delta t)]\Biggr\},
\label{f21aaa11}
\end{eqnarray}
where the functions ${\cal A}(\Delta\omega, \alpha_1, \alpha_2)$ and ${\cal I}(\Delta\omega, \Delta t)$ are defined in the~\ref{app2}. Expression~(\ref{f21aaa11}) displays the usual thermal Planck factor. Observe the presence of contributions that contain information about the temperature $\beta _{1}$ as well as terms comprising the knowledge on the temperature $\beta _{2}$. 

From the analysis carried out in the~\ref{app2}, one may easily take the limit of equal accelerations $\alpha _{1}=\alpha _{2}=\alpha$. For $\Delta t \gg \epsilon$, the associated transition rate is given by
%\begin{equation}
%\lim _{\Delta t \rightarrow \infty} \frac{dF_{21}(\Delta \omega ,\Delta t)}{d(\Delta t)}=-\frac{1}{16\pi ^{2}\alpha ^{2}}\int _{-\infty}^{\infty} d\psi \frac{e^{-i\Delta \omega \psi}}{\sinh\left(\frac{\psi}{2\alpha }-\frac{2i\epsilon}{\alpha }+\frac{\phi}{2} \right)\sinh\left(\frac{\psi}{2\alpha }-\frac{2i\epsilon}{\alpha}-\frac{\phi}{2} \right)}.
%\end{equation}
\beq
\lim_{\epsilon \to 0} \lim _{\Delta t \rightarrow \infty}
R_{21}(\Delta \omega ,\Delta t) = \frac{\sin (|\Delta \omega | \alpha \phi)}{2\pi \alpha \sinh\phi}
\left\lbrace \Theta (-\Delta \omega) \left[1+\frac{1}{e^{2\pi \alpha |\Delta \omega|}-1}\right] 
+ \Theta (\Delta \omega)\frac{1}{e^{2\pi \alpha \Delta \omega}-1}\right\rbrace .
\eeq
One obtains a similar expression for $R_{12}(\Delta \omega ,\Delta t)$ within the same asymptotic limit. 

Let us now evaluate the response functions $F_{12}(\Delta \omega ,\Delta t)$ and $F_{21}(\Delta \omega ,\Delta t)$ with the Gaussian form for the switching functions given by Eqs.~(\ref{gauss1}) and~(\ref{gauss2}). One has that
\bea
F_{21}(\Delta \omega ,\zeta) &=& -\frac{\sqrt{2 \pi } \zeta}{32\pi ^{2}\alpha _{1}^{2}}
e^{-[a_{-} \Delta\omega/(8 \alpha _1^2)](a_{-} \zeta ^2 \Delta\omega +8 i \alpha _1 \tau)}
\int _{-\infty} ^{\infty}d\psi 
\,e^{-i\Delta \omega (a_{+}/2\alpha _{1})\psi}\,e^{-\psi^2/(2 \zeta^2)}
\nn\\
&\times&\,\left[\sinh\Biggl(\frac{\psi}{2\alpha _{1}}-\frac{4i\epsilon}{a_{+}}+\frac{\phi}{2} \Biggr) 
\sinh\left(\frac{\psi}{2\alpha _{1}}-\frac{4i\epsilon}{a_{+}}-\frac{\phi}{2} \right)\right]^{-1}.
\eea
Proceeding as above, we singularize two situations of interest, namely $\zeta \lesssim \epsilon \ll 1$ and also $\zeta \gg \epsilon$. Within the prior conception one gets, for the former
\bea
F_{21}(\Delta \omega ,\zeta) &\approx& \frac{\zeta^2}{8\pi\alpha _{1}^{2}}
\,\frac{e^{-8 i \alpha _1 \tau[a_{-} \Delta\omega/(8 \alpha _1^2)]}}
{\cosh\phi - \cos\Bigl(\frac{8\epsilon}{a_{+}}\Bigr)}.
\eea
On the other hand, the case in which $\zeta \gg \epsilon$ can be handled as previously. One obtains
\bea
\hspace{-4mm}
F_{21}(\Delta \omega ,\zeta) &=& \frac{\sqrt{2 \pi } \zeta}{4\pi\alpha _{1}\sinh(\phi)}
e^{-[a_{-} \Delta\omega/(8 \alpha _1^2)](a_{-} \zeta ^2 \Delta\omega +8 i \alpha _1 \tau)}
\nn\\
&\times&\,
\left\{\Theta(\Delta\omega)\,\exp{\left(\frac{2\alpha_1^2{\cal D}(\Delta\omega)}{a_{+}^2\zeta^2}\right)}
\left[e^{-\frac{4 a_{-}\Delta\omega\epsilon}
{a_{+}}} \sin \left(\frac{a_{+} \Delta\omega \phi}{2}\right)\,
\frac{1}{e^{\pi a_{+} \Delta\omega }-1}\right]
\right.
\nn\\
&+&\, \left. \Theta(-\Delta\omega)\,\exp{\left(\frac{2\alpha_1^2{\cal D}(|\Delta\omega|)}{a_{+}^2\zeta^2}\right)}
\left[e^{-\frac{4 a_{-}|\Delta\omega|\epsilon}
{a_{+}}} \sin \left(\frac{a_{+} |\Delta\omega| \phi}{2}\right)
\left(1 + \frac{1}{e^{\pi a_{+} |\Delta\omega| } - 1}\right)\right]\right\}.
\eea
It is easy to see that cross-correlation terms strongly depend on the configuration chosen for the switching functions. In particular, notice that the thermal terms are associated with a temperature $1/(\pi a_{+})$. Possibly the origin of such distinguished patterns stems from the fact that Gaussian switching attenuates the amount of local noise undergone by particle detectors and as such stimulates the prevalence of nonlocal terms over local noise contributions~\cite{emm}. Clearly this will have an important impact over the probability transition to the entangled states $\left| \Psi^{\pm} \right\rangle$. This behavior will be fully addressed in the next section. In any case we note that, for equal accelerations $\alpha_1 = \alpha_2 = \alpha$ and $\zeta \gg \epsilon$, the leading contribution to the associated transition rate is given by ($\Delta t = \sqrt{2 \pi } \zeta$)
\beq
\lim_{\epsilon \to 0} \lim _{\Delta t \rightarrow \infty}
R_{21}(\Delta \omega ,\Delta t) = \frac{\sin (|\Delta \omega | \alpha \phi)}{4\pi \alpha \sinh\phi}
\left\lbrace \Theta (-\Delta \omega) \left[1+\frac{1}{e^{2\pi \alpha |\Delta \omega|}-1}\right] 
+ \Theta (\Delta \omega)\frac{1}{e^{2\pi \alpha \Delta \omega}-1}\right\rbrace,
\eeq
which closely resembles the result employing sharp switching functions. Same considerations apply to the response function $F_{12}$; here again $F_{21}=F_{12}^{*}$.

We are now ready to specialize the considerations of this section to particular transitions involving entangled states.

\section{Radiative processes of entangled states}

\begin{figure}[t]
\begin{center}
\includegraphics[scale=0.34]{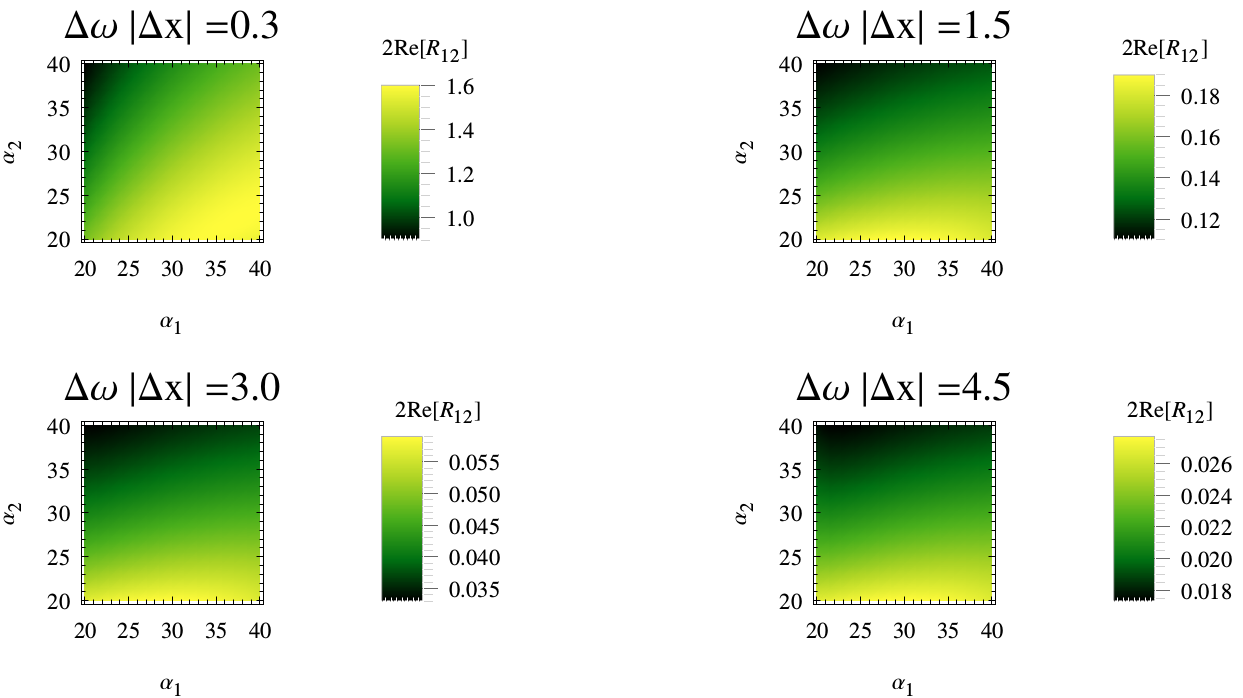}\\
\caption{The quantity $\textrm{Re}[R_{12}]_{\left| gg\right\rangle \rightarrow\left| \Psi ^{+}\right\rangle}$ (with the sharp switching function) as a function of the inverse accelerations $\alpha_1, \alpha_{2}$ for $\Delta \omega \Delta t = 1.2$, $\Delta \omega \epsilon = 3.0 \times 10^{-2}$ and different values of $\Delta\omega |\Delta \mathbf{x}|$. We also considered a symmetric proper time interval about the origin, which amounts to take $T=0$. The quantity $\Delta \omega |\Delta \mathbf{x}|$ serves as control parameter in the study of transition probabilities concerning entangled states. All physical quantities are given in terms of the natural units associated with the specific transition $\left| gg\right\rangle \rightarrow\left| \Psi ^{+}\right\rangle $. Therefore, in this case $\xi_1$, $\xi_2 $ and $\alpha_1, \alpha_2$ are measured in units of $\lambda$, and $\omega_0$ is given in units of $2\pi\lambda^{-1}$, where $\lambda = 2\pi/\omega_0$. Moreover, $\textrm{Re}[R_{12}]$ is measured in units of $\lambda^{-1}$.}
\label{primera}
\end{center}
\end{figure}

\begin{figure}[h]
\begin{center}
\includegraphics[scale=0.45]{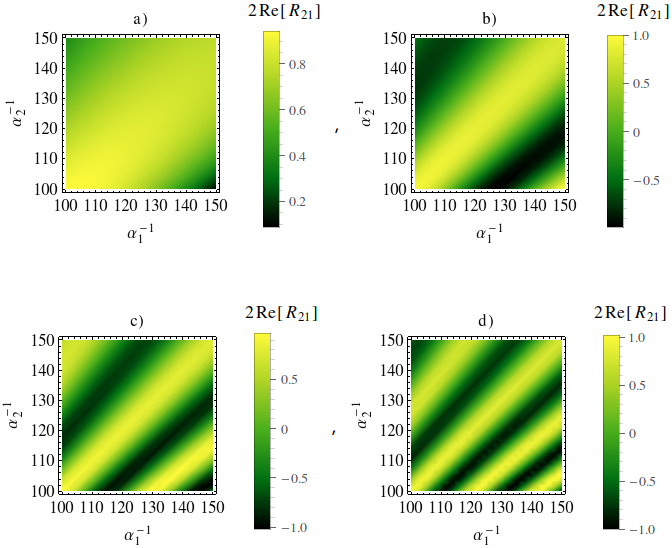}\\
\caption{The quantity $\textrm{Re}[R_{21}]_{\left| gg\right\rangle \rightarrow\left| \Psi ^{+}\right\rangle}$ (with the sharp switching function) as a function of the inverse accelerations $\alpha_1, \alpha_{2}$ for different values of $\Delta\omega\Delta t$. We consider a fixed value $\Delta \omega |\Delta \mathbf{x}|=0.3$ and $\Delta \omega \epsilon =3.0\times 10^{-2}$ in the four cases. In a) 
$\Delta\omega\Delta t = 3.0$, b) $\Delta\omega\Delta t = 12.0$, c) $\Delta\omega\Delta t = 21.0$, 
d) $\Delta\omega\Delta t = 30.0$. As in the previous figure we considered $T=0$. Here it is clear that maximum values show up in other regions besides the region in which $\alpha _{1}=\alpha _{2}$. The inverse accelerations $\alpha_1, \alpha_2$ are measured in units of $\lambda$ and $\textrm{Re}[R_{12}]$ is measured in units of $\lambda^{-1}$ (see caption of Fig.\ref{primera} for a definition of $\lambda$).}
\label{grandesacc}
\end{center}
\end{figure}

\begin{figure}[h]
\begin{center}
\includegraphics[scale=0.45]{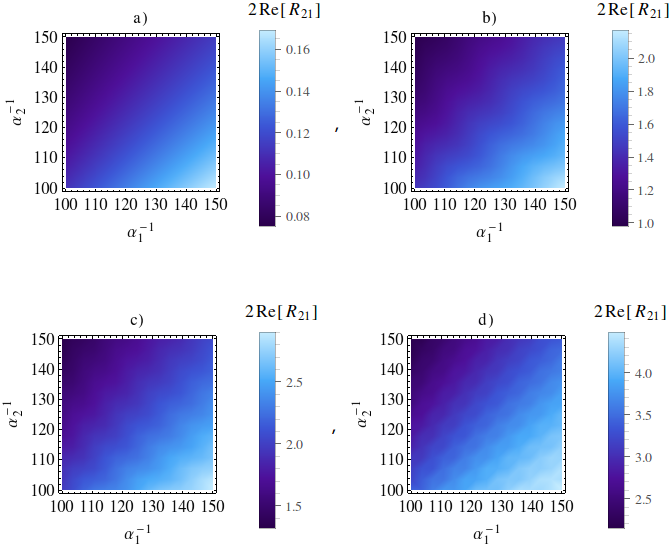}\\
\caption{The quantity $\textrm{Re}[R_{21}]_{\left| gg\right\rangle \rightarrow\left| \Psi ^{+}\right\rangle}$ (with the Gaussian switching function) as a function of the inverse accelerations $\alpha_1, \alpha_{2}$ for different values of $\Delta\omega\zeta$ where $\zeta \lesssim \epsilon \ll 1$. We consider a fixed value $\Delta \omega |\Delta \mathbf{x}|=0.3$ in the four cases. In a) $\Delta\omega \epsilon = 3.0 \times 10^{-2}$; $\Delta\omega \zeta = 2.7\times 10^{-2}$, b) $\Delta\omega \epsilon = 3.9\times 10^{-1}$; $\Delta\omega \zeta = 35.1\times 10^{-2}$, c) $\Delta\omega \epsilon = 5.4\times 10^{-1}$; $\Delta\omega \zeta = 48.6\times 10^{-2}$ and d) $\Delta\omega \epsilon = 0.9$; $\Delta\omega \zeta = 8.1\times 10^{-1}$. We also take $\tau = (0.9\sqrt{2\pi}/2) \epsilon $. Here we do not have maximum values in the region in which $\alpha _{1}=\alpha _{2}$. The inverse accelerations $\alpha_1, \alpha_2$, as well as $\zeta$ and $\epsilon$, are measured in units of $\lambda$ and $\textrm{Re}[R_{12}]$ is measured in units of $\lambda^{-1}$ (see caption of Fig.\ref{primera} for a definition of $\lambda$).}
\label{r12_gaussianp}
\end{center}
\end{figure}

\begin{figure}[h]
\begin{center}
\includegraphics[scale=0.45]{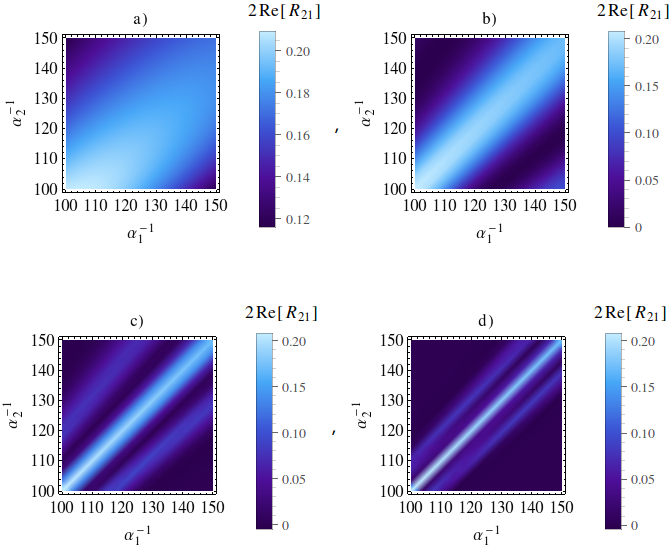}\\
\caption{The quantity $\textrm{Re}[R_{21}]_{\left| gg\right\rangle \rightarrow\left| \Psi ^{+}\right\rangle}$ (with the Gaussian switching function) as a function of the inverse accelerations $\alpha_1, \alpha_{2}$ for different values of $\Delta\omega\sqrt{2\pi} \zeta$ where $\zeta \gg \epsilon $. We consider a fixed value $\Delta \omega |\Delta \mathbf{x}|=0.3$ in the four cases and $\tau$ is the same as in the previous figure. In a) $\Delta\omega \sqrt{2\pi} \epsilon = 0.3\times 10^{-2}$; $\Delta\omega \sqrt{2\pi} \zeta = 3.0$, b) $\Delta\omega \sqrt{2\pi} \epsilon = 1.2\times 10^{-2}$; $\Delta\omega \sqrt{2\pi} \zeta = 12.0$, c) $\Delta\omega \sqrt{2\pi} \epsilon = 3.6\times 10^{-2}$; $\Delta\omega \sqrt{2\pi} \zeta = 36.0$ and d) $\Delta\omega \sqrt{2\pi} \epsilon = 7.2 \times 10^{-2}$; $\Delta\omega \sqrt{2\pi} \zeta = 72.0$. Here we have maximum values in the region in which $\alpha _{1}=\alpha _{2}$ and $\alpha _{1}\neq\alpha _{2}$. The inverse accelerations $\alpha_1, \alpha_2$ are measured in units of $\lambda$ and $\textrm{Re}[R_{12}]$ is measured in units of $\lambda^{-1}$ (see caption of Fig.\ref{primera} for a definition of $\lambda$).}
\label{r12_gaussianm1}
\end{center}
\end{figure}

This section encompasses the study of the transition rate to entangled states. We focus attention on the particular transition $\left| gg\right\rangle \rightarrow \left| \Psi ^{+}\right\rangle$, but a similar analysis can be implemented for the entangled state  $\left| \Psi^{-} \right\rangle$. The corresponding matrix elements of this transition are given by
\begin{equation}
m_{gs}^{(1)} = m_{gs}^{(2)} = 1/\sqrt{2},
\label{matrix-elements}
\end{equation}
and the gap energy is $\Delta\omega = \omega _{0}$. In Eq.~(\ref{matrix-elements}) the label $s$ refers to the fact that $\left| \Psi ^{+}\right\rangle$ is a symmetric entangled state. We are particularly interested in the contributions generated by cross correlations between the atoms. Therefore we define the cross contribution for the total transition rate as $\textrm{Re}[R_{12}]_{\left| gg\right\rangle \rightarrow\left| \Psi ^{+}\right\rangle}$ which is properly evaluated in~\ref{app2}, see the expression~(\ref{r12app}). The behavior of such a quantity as a function of the inverse accelerations $\alpha_1, \alpha_{2}$ is depicted in the Fig.~\ref{primera}, for a fixed small time interval and different spatial separations $|\Delta \mathbf{x}|$. One plainly observes the occurrence of maximum values for $\textrm{Re}[R_{12}]$ for specific values of the accelerations. This result primarily demonstrates how $|\Delta \mathbf{x}|$ can be employed as a control parameter for the transition to the entangled state 
$\left| \Psi ^{+}\right\rangle$ from the ground state. As expected, a large value of $|\Delta \mathbf{x}|$ corresponds to a significant reduction on the magnitude of the cross contribution. In turn, the results for different time intervals are summarized in Fig.~\ref{grandesacc}. Observe that the mutual influences of the atoms in this situation implies in a rather distinct interference pattern in comparison with the previous case. In addition, not only the region $\alpha _{1} = \alpha _{2}$ in the plot produces sensible contributions to the transition rate, but we note the appearance of other regions in the plot that also provide important contributions.

Let us now discuss the outcomes engendered by the Gaussian form of the switching functions, given by Eqs.~(\ref{gauss1}) and~(\ref{gauss2}). First consider the regime in which $\zeta \lesssim \epsilon \ll 1$. Such a situation is depicted in Fig.~\ref{r12_gaussianp}. In this case, the short time intervals do not allow the formation of a interference pattern. Observe that the maximum values are not located in the $\alpha _{1}=\alpha _{2}$ region. In comparison with the analogous situation presented in Fig.~\ref{grandesacc} one can notice a serie of maximum values showing up in a orthogonal form. On the other hand, the situation is drastically modified when one considers instead the regime $\zeta \gg \epsilon$. The results are summarized in Fig.~\ref{r12_gaussianm1}. In this case, we were able to explore large time intervals where we have the formation and disappearance of a interference pattern. As one allows for larger values of $\zeta$, one observes that the maximum values have a tendency to cluster around the line $\alpha _{1}=\alpha _{2}$. This implies that for large times only atoms in the same world line in $(t,z)$-plane will have a large crossed contribution. This represents a substantial divergence from the behaviour observed in Fig.~\ref{grandesacc}. The previous assertion regarding the profound dependence of cross-correlation terms on the profile of the switching functions is manifest in this way. For equal accelerations both switching functions present the same qualitative behavior in this asymptotic regime, namely maximum values located along the line $\alpha _{1}=\alpha _{2}$. Moreover, for both cases large accelerations (temperatures) indicate reduction in the cross contributions. 

To conclude this Section, let us briefly discuss the total transition rate~(\ref{rtotal1}) within the asymptotic time interval regime and for small distances between the atoms. We also consider the case of equal accelerations. In this situation one can express the total transition rate as follows
\begin{equation}
\mathcal{R} _{\left| gg \right\rangle \rightarrow \left| \Psi ^{+} \right\rangle}= R_{11}f(\omega_0\alpha \phi),
\label{ratet11}
\end{equation}
where $f(x)=2\left(1 + \sin x/x\right)$. This function quantifies the influence of the crossed response functions on the entanglement between atoms, for asymptotic time intervals. Some special values are given by ($n$ is a positive integer)
\begin{equation}
f[(2n+1)\pi /2]=2\left(1+\frac{2(-1)^{n}}{(2n+1)\pi}\right).
\end{equation}
A closer inspection in the behaviour of this function reveals that there is a great oscillatory regime for large accelerations and small distances between the atoms. Since the atoms have the same $z$ coordinate, for $\phi \ll 1/\Delta \omega \alpha$ cross correlations are more important for the rate~(\ref{ratet11}) in comparison with cases in which the
$\Delta\omega|\Delta \mathbf{x}|$ becomes larger. Therefore, the crossed response functions generate a constructive interference when the atoms are near each other in space. In turn, these interference terms vanish for large spatial separations between the atoms. Similar conclusions were reported in Refs.~\cite{hu2,lenz1,ariase1,gabriel1,gabriel3}.

\section{Mean life of entangled states}

\begin{figure}[h]
\begin{center}
\includegraphics[scale=0.45]{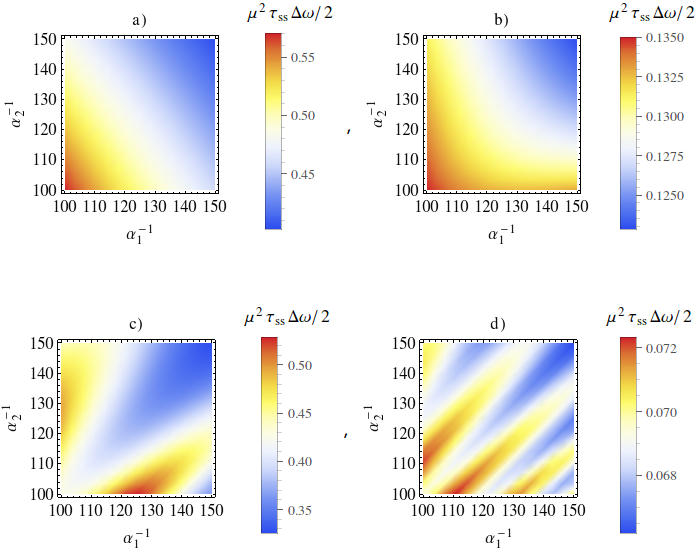}\\
\caption{The quantity $\mu^{2}|\Delta\omega |\tau _{ss}/2$ (with the sharp switching function) as a function of the accelerations of the atoms for different values of $\Delta\omega\Delta t$. We consider the fixed values $\Delta \omega |\Delta \mathbf{x}|=0.3$ and $\Delta \omega \epsilon =3.0\times 10^{-2}$ in the four cases. In a) $\Delta\omega\Delta t = 0.3$, b) $\Delta\omega\Delta t = 3.0$, c) $\Delta\omega\Delta t = 12.0 $ and d) $\Delta\omega\Delta t = 30.0 $. Again we considered a symmetric proper time interval about the origin, $T=0$. We have the presence of interference effects that softens the decay of the entangled state $\left| \Psi ^{+}\right\rangle$. The inverse accelerations $\alpha_1, \alpha_2$ are measured in units of $\lambda$ (see caption of Fig.\ref{primera} for a definition of $\lambda$).}
\label{meanlifetss1}
\end{center}
\end{figure}

\begin{figure}[h]
\begin{center}
\includegraphics[scale=0.45]{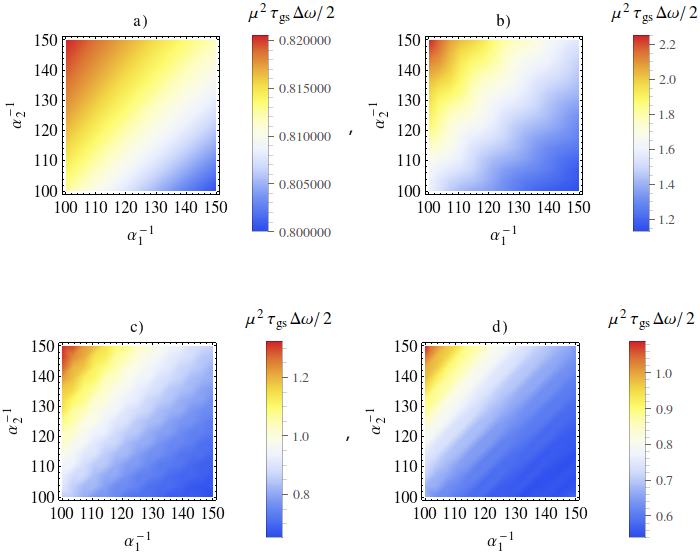}\\
\caption{The quantity $\mu^{2}|\Delta\omega |\tau _{gs}/2$ (with the Gaussian switching function) as a function of the accelerations of the atoms for different values of $\Delta\omega\zeta$ where $\zeta \lesssim \epsilon \ll 1$. We consider a fixed value $\Delta \omega |\Delta \mathbf{x}|=0.3$ in the four cases and take $\tau$ as in Fig.~\ref{r12_gaussianp}. In a) $\Delta\omega \epsilon = 3.0\times 10^{-2}$; $\Delta\omega \zeta = 2.7\times 10^{-2}$, b) $\Delta\omega \epsilon = 4.5\times 10^{-1}$; $\Delta\omega \zeta = 40.5\times 10^{-2}$, c) $\Delta\omega \epsilon = 0.9$; $\Delta\omega \zeta = 8.1\times 10^{-1}$ and d) $\Delta\omega \epsilon = 1.2$; $\Delta\omega \zeta = 10.8\times 10^{-1}$. The inverse accelerations $\alpha_1, \alpha_2$ are measured in units of $\lambda$ (see caption of Fig.\ref{primera} for a definition of $\lambda$).}
\label{meanlifetgs1}
\end{center}
\end{figure}

\begin{figure}[h]
\begin{center}
\includegraphics[scale=0.45]{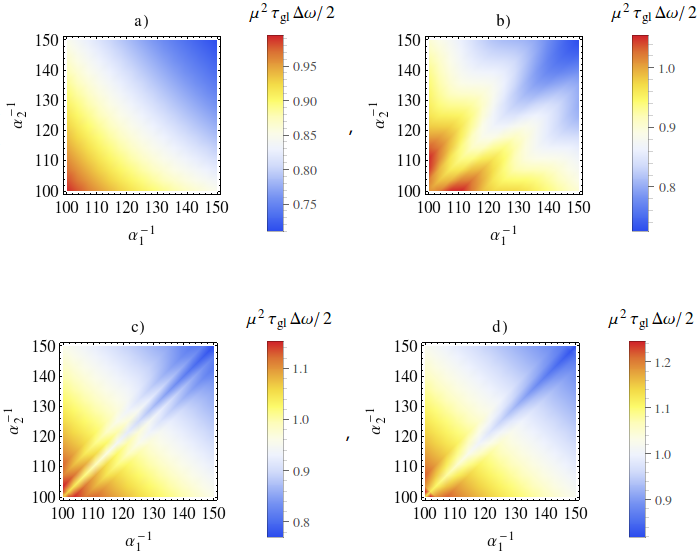}\\
\caption{The quantity $\mu^{2}|\Delta\omega |\tau _{gl}/2$ (with the Gaussian switching function) as a function of the accelerations of the atoms for different values of $\Delta\omega\zeta$ where $\zeta \gg \epsilon$. We consider a fixed value $\Delta \omega |\Delta \mathbf{x}|=0.3$ in the four cases and take 
$\tau$ as in Fig.~\ref{r12_gaussianp}. In a) $\Delta\omega \sqrt{2\pi} \epsilon = 0.3\times 10^{-2}$; $\Delta\omega \sqrt{2\pi} \zeta = 3.0$, b) $\Delta\omega \sqrt{2\pi} \epsilon = 0.3\times 10^{-1}$; $\Delta\omega \sqrt{2\pi} \zeta = 30.0$, c) $\Delta\omega \sqrt{2\pi} \epsilon = 10.8\times 10^{-2}$; $\Delta\omega \sqrt{2\pi} \zeta = 108.0$ and d) $\Delta\omega \sqrt{2\pi} \epsilon = 18.9 \times 10^{-2}$; $\Delta\omega \sqrt{2\pi} \zeta = 189.0$. Here we have the presence of interference effects that provide more stability for the entangled state $\left| \Psi ^{+}\right\rangle$ for atoms with $\alpha _{1}\neq \alpha _{2}$. The inverse accelerations $\alpha_1, \alpha_2$ are measured in units of $\lambda$ (see caption of Fig.\ref{primera} for a definition of $\lambda$).}
\label{meanlifetgl1}
\end{center}
\end{figure}

So far we have studied the quantum mechanical transition probability to entangled states through the excitation of the collective ground state $\left| gg \right\rangle$. We have demonstrated the possibility that, for uniformly accelerated atoms, the interaction with a common quantum field can be an essential ingredient in assessing the probability transition to the entangled state $\left| \Psi ^{+}\right\rangle$. On the other hand, such an interaction can also induce effects connected with entanglement degradation. Hence a natural question that emerges is whether such entangled states persist for long time intervals. A possible measurement of the decay of entangled states is given by the mean life of such states. The mean life for the state $|\omega'\rangle$ is defined as
\begin{equation}
\tau _{\left| \omega ' \right\rangle \rightarrow \left| \omega \right\rangle}(\Delta \omega ,\Delta t)=
[\mathcal{R} _{\left| \omega ' \right\rangle \rightarrow \left| \omega \right\rangle}(\Delta \omega ,\Delta t)]^{-1}.
\label{tiempom}
\end{equation}
In general, in order to address the degradation of entanglement, one should study the transition $\left| \Psi ^{+}\right\rangle \rightarrow\left| \omega \right\rangle$, where $|\omega\rangle$ is a generic separable state. Here we are interested in investigating the stability of the entangled state under spontaneous emission processes, confining our discussions to the case associated with the transition $\left| \Psi ^{+}\right\rangle \rightarrow \left| gg \right\rangle$. Notwithstanding, we strongly remark that, for a more general analysis, one should also study the transition of the entangled state to other separable states in order to have a complete knowledge of the timescale for the nonlocal decoherence of the state 
$\left| \Psi ^{+}\right\rangle$.

The corresponding matrix elements of the transition $\left| \Psi ^{+}\right\rangle \rightarrow \left| gg \right\rangle$ are given by Eq.~(\ref{matrix-elements}), and the gap energy now is given by $\Delta\omega = -\omega _{0}$. Hence the expression (\ref{tiempom}) becomes
\beq
\tau_{\left| \Psi ^{+} \right\rangle \rightarrow \left| gg \right\rangle} =
\frac{2}{\mu^{2}}\Bigl\{R_{11}(-\omega_0 ,\Delta t)+R_{22}(-\omega_0 ,\Delta t)
+ 2\textrm{Re}[R_{12}(-\omega_0 ,\Delta t)] \Bigr\}^{-1}.
\eeq
Let us define $\tau _{ss}$, $\tau _{gs}$ and $\tau _{gl}$, the mean life of the entangled states when we have a sharp switching function, a Gaussian switching function when $\zeta \lesssim \epsilon \ll 1$ and a Gaussian switching function when $\zeta \gg \epsilon$, respectively. The behavior of the mean life $\tau _{ss}$ as a function of the accelerations for different time intervals and with the condition $|\Delta\omega||\Delta \mathbf{x}| \ll 1$ is depicted in the Fig.~\ref{meanlifetss1}. Note that such a function falls off quickly with the acceleration. This result has a clear-cut meaning: the entangled state $\left| \Psi ^{+}\right\rangle$ rapidly decays for large observational times and also for sufficiently large accelerations. We note the emergence of an oscillatory regime. Observe that the mean life of the entangled state $\left| \Psi ^{+}\right\rangle$ for large $\Delta t$ displays maximum values not only in the line $\alpha _{1} = \alpha _{2}$ but also in other regions.

The behavior of the mean life $\tau _{gs}$ as a function of the accelerations for different time intervals and with the condition $\zeta \lesssim \epsilon \ll 1$ is depicted in the Fig.~\ref{meanlifetgs1}. In this case we observe a relative stability for the entangled state $\left| \Psi ^{+}\right\rangle$. Moreover, notice that the maximum values are higher in comparison with the analogous situation represented in Fig.~\ref{meanlifetss1}. This suggests that for small time intervals a Gaussian switching function exhibits a stronger attenuation in the decay of the entangled state as compared with a sharp switching function. This distinction may be envisaged as the result of a very short switching time, which for the case of a sharp switching function produces a large disturbance in the system.

One finds illustrated in the Fig.~\ref{meanlifetgl1} the behavior of the mean life $\tau _{gl}$ as a function of the accelerations for different time intervals and with the condition $\zeta \gg \epsilon$. In this case for large observational times we notice the emergence of an oscillatory regime as observed in the Fig.~\ref{meanlifetss1}. The maximum values increases with $\zeta$ and are higher than the corresponding values for the sharp switching function. This again indicates that the stability of the entangled state $\left| \Psi ^{+}\right\rangle$ is much more expressive in the case of smooth switching. Furthermore, the entangled state $\left| \Psi ^{+}\right\rangle$ strongly decays for sufficiently large accelerations. In addition, in contrast with the sharp-switching case, the maximum values get bunched up along the line $\alpha_{1} = \alpha_{2}$ for larger observational time intervals. For equal accelerations and large times we observe again a similar qualitative behavior for both switching functions. In addition, we have showed that we can obtain a relative entanglement stability for atoms following different world lines, i.e. subjected to different temperatures. It is not necessary to have strictly close atoms in space in order to have entanglement stability for an adequate choice of observational time interval.

\section{Summary and outlook}

In this paper we were interested in uncovering certain aspects related to radiative processes of entangled states. Being more specific, we have studied two identical uniformly accelerated two-level atoms weakly coupled with a massless scalar field prepared in the Minkowski vacuum state. We have shown the existence of a non-vanishing transition probability to the symmetric maximally entangled state for atoms initially prepared in the ground state. We also found that the associated response function contains terms related to cross correlations between the atoms mediated by the field. Since the atoms move along different world lines, such crossed terms present thermal contributions with different temperatures. The crossed terms of the response function are modulated by an oscillating function. In addition, such contributions are accompanied by a gray body factor whose arguments have information on the accelerations and the energy gap. The appearance of the aforementioned gray body factor may be understood in the sense of Ref.~\cite{hu3}, in which the authors demonstrate the emergence of a thermal noise produced by the fluctuations of the fields and field correlations between the two trajectories. Moreover, we also presented an expression for the mean life of the entangled state $\left| \Psi ^{+}\right\rangle$ by taking into account only the case associated with the transition $\left| \Psi ^{+}\right\rangle \rightarrow \left| gg \right\rangle$. In general, we found that atoms with same acceleration will be less correlated for an increasing $|\Delta \mathbf{x}|$. 

We have considered different settings for the coupling between atoms and fields. Concerning the limit of sharp switching of the interactions, we have shown that, for large observational times, contributions coming from cross correlations present higher values along different lines in the plots exposed above. On the other hand, in direct contrast with this result, as the observational time increases, the maximum allowed values for the cross contributions cluster more and more along the line $\alpha_{1} = \alpha_{2}$ for the Gaussian switching function. This has at least two important consequences. One is that for large times, only atoms with similar accelerations will have a large cross contribution for the Gaussian case. Indeed, the only case in which both switching functions displays the same patterns is for equal accelerations. In turn, the stability of the entangled state $\left| \Psi ^{+}\right\rangle$ is more pronounced for the smooth switching as compared to the sudden switching, for a suitable choice of parameters. Such an assertion is in line with the inferences claimed by the authors of Ref.~\cite{emm} concerning entanglement harvesting. Furthermore, the cross correlations are very sensitive to the specifications of how the interaction between atoms and fields is turned on and off.  On the other hand, it is known that, close to the event horizon, the Schwarzschild metric approximately takes the form of a Rindler line element. Hence for atoms placed at points that are close enough to the event horizon, in principle the outcomes discussed in this work can be extended to Schwarzschild black holes. Even though our simple analysis do not cover several important aspects concerning entanglement between two-level atoms, one cannot avoid to engage in a wishful speculation that, depending on the conditions imposed, our results could be of some relevance in estimating the effects of loss of entanglement as the atoms approach the event horizon. Indeed, the investigation of Ref.~\cite{Henderson:17} seems to indicate that black holes in fact hinder entanglement harvesting. In a certain sense, this agrees with the outcomes of Ref.~\cite{sch}.

It has been reported that the entanglement harvesting by two atoms (detectors) with anti-parallel acceleration undergoes an enhancement, whereas for two atoms undergoing parallel acceleration one finds instead a degradation of entanglement harvesting~\cite{Salton:15}. In such a work, the authors consider a vanishing orthogonal separation between the atoms, but they introduce a non-zero distance of closest approach between the atoms (as measured by an inertial observer at fixed $z$) in the coordinate $z$ (in our notation). In addition, they keep the atoms in distinguished Rindler wedges. In our approach we keep the atoms atoms within the same Rindler wedge, but with different proper accelerations, and with a non-vanishing orthogonal spatial separation $|\Delta \mathbf{x}|$. For large 
$|\Delta \mathbf{x}|$ or sufficiently high proper accelerations, cross-correlations will be suppressed. So in principle one could take this as an evidence that for large orthogonal separation, or large accelerations, entanglement harvesting could be degraded, corroborating the outcomes in Ref.~\cite{Salton:15} concerning the situation of atoms accelerating in the same direction at the same rate. In turn, the case with equal proper accelerations (but non-vanishing $|\Delta \mathbf{x}|$) in the present analysis seems to indicate a possible increase in the entanglement harvesting since cross-correlations between the atoms exhibit maximum values along the line $\alpha_1 = \alpha_2$. This would be similar to the results found for the anti-parallel acceleration case treated in Ref.~\cite{Salton:15}. In any event, since the settings and purposes in the present case and in Ref.~\cite{Salton:15} are essentially distinct, it is not immediate clear how each result directly compare with each other, so such inferences should be perceived with adequate caution. All the aforementioned differences could have an impact in the final conclusions. A careful treatment that could suitably incorporate both situations would be most welcome, but would need other refined tools that were not employed in the present analysis.  

The present work comprises a wide variety of possible generalizations. It is clear that a more detailed analysis of entanglement formation and degradation requires more sophisticated techniques such as the evaluation of entanglement monotones (concurrence, negativity, etc.). Furthermore, one could also consider a more thorough treatment as for instance the one provided by the master equation approach. This would enable more general conclusions concerning quantum entanglement between atoms coupled with quantum fields. In this direction, we recall Refs.~\cite{Benatti:04,hu:11} which study the entanglement generation in Rindler and Schwarzschild space-times in a framework of open quantum systems. Similarly, one of the authors has recently put forward an elaborated investigation of entanglement in a Kerr space-time~\cite{gsm:18}. In turn, in Ref.~\cite{Zarro:2017} the authors investigated how entanglement dynamics between two two-level atoms is disturbed in a disordered medium. Hence a possible extension (and perhaps a more complete exploration) of the results presented here can be fully probed by the usage of such an approach. On the other hand, one could assess the impact on our conclusions regarding the presence of material boundaries, employing similar configurations as the ones considered in~\cite{ariase1}. One could also investigate the case of spatially extended, uniformly accelerated Unruh-DeWitt detectors~\cite{emm,ivette}.  Such subjects are under investigation by the authors.

\section*{Acknowledgements}

This work was partially supported by Conselho Nacional de Desenvolvimento Cient\'ifico e Tecnol\'ogico -- CNPq and Coordena\c{c}\~ao de Aperfei\c{c}oamento de Pessoal de N\'ivel Superior -- CAPES (Brazilian agencies).

\appendix

\section{Explicit calculation of individual response functions}
\label{app1}

\begin{figure}[h]
\begin{center}
\includegraphics[scale=0.68]{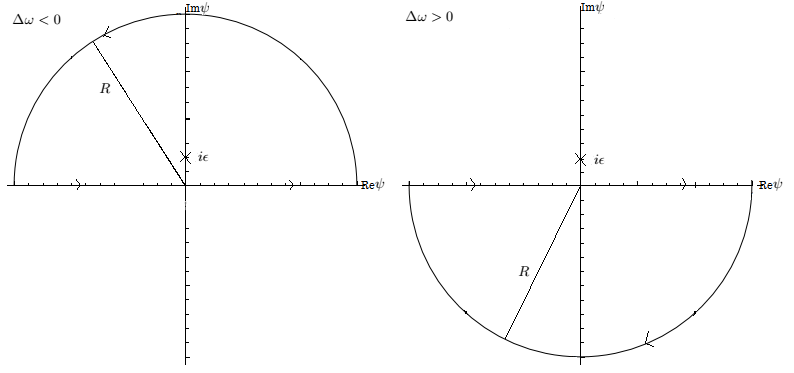}\\
\caption{Contour used to perform the integral of $F_{11}(\Delta \omega ,\Delta t)$ and $F_{22}(\Delta \omega ,\Delta t)$.}
\label{contorno1}
\end{center}
\end{figure}

In this Appendix we concisely perform the evaluation of the individual contributions of the atoms to the total response function in the case of sharp switching defined by Eqs.~(\ref{sharp1}) and~(\ref{sharp2}). We only consider the case of the response function $F_{11}$ in detail since, as remarked in the text, all the results associated with the response function $ F_{22}$ can be obtained from $F_{11}$ by performing the replacement $\Delta\omega \to (\alpha_{2}/\alpha_{1})\Delta\omega$. 

In order to study the contribution (\ref{f11integral}), we perform the Fourier transform with the help of contour-integration methods. From the expression (\ref{g11ii}) one notes the existence of second order poles of the form $\psi_{n} =2i\epsilon + 2\pi i\alpha _{1} n$, where $n$ is an integer. One must treat separately the cases of $n\neq 0$ and $n=0$. For $\Delta \omega < 0$ we make use of a semicircle of radius $R$ that we close on the upper-half $\textrm{Im}[\psi] > 0$ plane. This contour encloses the poles for $n\geq 0 $ and runs in an anticlockwise direction. For $\Delta \omega > 0$ we close the contour in a semicircle of radius $R$ in the lower-half $\textrm{Im}[\psi] < 0$ plane. Now, this contour encloses the poles for $n < 0$ and runs in the clockwise direction (see Fig.~\ref{contorno1}). We consider the limit $R\rightarrow \infty$ such that the contribution from the arcs will vanish by the Jordan's lemma. We obtain, for the atom $1$
\begin{eqnarray}
 F_{11}(\Delta \omega ,\Delta t) &=& \frac{\Delta t}{2\pi ^{2}}
\left\lbrace e^{-2\epsilon|\Delta \omega|}\pi |\Delta \omega |\Theta (-\Delta \omega)\left[1 + \frac{1}
{e^{2\pi \alpha _{1}|\Delta \omega |}-1}\right] \right.
\nonumber\\
&+&\, \left. \frac{e^{2\epsilon\Delta \omega}\pi\Delta \omega\Theta (\Delta \omega)}{e^{2\pi \alpha _{1}\Delta \omega }-1}
+ {\cal P}_{1}(\Delta\omega, \Delta t, \epsilon) + \Re\left[{\cal J}_{1}(\Delta\omega, \Delta t, \epsilon)\right]\right\rbrace
\nonumber \\
&+&\,  \frac{1}{2\pi ^{2}}\biggl\{{\cal P}_{2}(\Delta\omega, \Delta t, \epsilon) 
+ \Re\left[{\cal J}_{2}(\Delta\omega, \Delta t, \epsilon)\right]\biggr\}.
\end{eqnarray}
This is the expression~(\ref{f11tiempofinito}). We have introduced the following quantities
\bea
\hspace{-8mm}
{\cal P}_{1}(\Delta\omega, \Delta t, \epsilon) &=& -\frac{\pi  \left| \Delta\omega \right|\,e^{2\epsilon\Delta\omega}}{2}  
+\frac{\Delta t \cos (\Delta t \Delta\omega ) + 2 \epsilon  \sin (\Delta t \Delta\omega )}{\Delta t^2+4 \epsilon ^2}
\nn\\
&-&\,2 \Re\left[| \Delta\omega|  e^{i \Delta t \Delta\omega} \textrm{Ci}\Bigl((\Delta t+2 i \epsilon ) | \Delta\omega| \Bigr) 
\sin\Bigl(|\Delta\omega| (\Delta t + 2 i \epsilon)\Bigr)\right]
\nn\\
&+&\,2\Im\left[i \Delta\omega  e^{-i \Delta t \Delta\omega } \textrm{Ci}\Bigl((\Delta t - 2 i \epsilon )| \Delta\omega| \Bigr) 
\cos\Bigl(|\Delta\omega|(\Delta t - 2 i \epsilon)\Bigr)\right]
\nn\\
&+&\,2 \Re\left[| \Delta\omega|  e^{i \Delta t \Delta\omega } \textrm{Si}\Bigl((\Delta t+2 i \epsilon ) | \Delta\omega| \Bigr) 
\cos\Bigl(|\Delta\omega| (\Delta t + 2 i \epsilon)\Bigr)\right]
\nn\\
&+&\,2\Im\left[i \Delta\omega  e^{-i \Delta t \Delta\omega } \textrm{Si}\Bigl((\Delta t - 2 i \epsilon )| \Delta\omega| \Bigr) 
\sin\Bigl(|\Delta\omega|(\Delta t - 2 i \epsilon)\Bigr)\right],
\eea
\bea
{\cal P}_{2}(\Delta\omega, \Delta t, \epsilon) &=&
-\left[\frac{\Delta t^2+ 4 \epsilon ^2 - 4 \epsilon ^2 \cos (\Delta t \Delta\omega ) + 2 \Delta t \epsilon  \sin (\Delta t \Delta\omega)}{\Delta t^2 + 4 \epsilon ^2} \right]
\nn\\
&+&\, \frac{e^{2 \Delta\omega \epsilon} (2 \Delta\omega  \epsilon +1)}{2} 
\Bigl\{\textrm{Ei}[(i\Delta t - 2 \epsilon) \Delta\omega] 
\nn\\
&+&\, \textrm{Ei}[-(i \Delta t + 2 \epsilon)\Delta\omega]
-2 \textrm{Ei}(-2 \epsilon \Delta\omega)\Bigr\},
\eea
\beq
{\cal J}_{1}(\Delta\omega, \Delta t, \epsilon) = \int _{\Delta t}^{\infty} d\psi e^{-i \Delta \omega \psi}
\left(\frac{1/(2\alpha _{1})^{2}}{\sinh ^{2}\left(\frac{\psi- 2i\epsilon}{2\alpha _{1}}\right)}-\frac{1}{(\psi-2i\epsilon)^{2}}\right),
\eeq
and finally
\beq
{\cal J}_{2}(\Delta\omega, \Delta t, \epsilon) = \int _{0}^{\Delta t} d\psi \psi\,e^{-i \Delta \omega \psi}
\left(\frac{1/(2\alpha _{1})^{2}}{\sinh ^{2}\left(\frac{\psi- 2i\epsilon}{2\alpha _{1}}\right)}-\frac{1}{(\psi-2i\epsilon)^{2}}\right).
\eeq
In the above expressions, $\textrm{Si}(z)$ is the sine integral function, $\textrm{Ci}(z)$ is the cosine integral function and $\textrm{Ei}(z)$ is the exponential integral function. Their standard definitions as well as important properties satisfied by them can be found in the Ref.~\cite{series2}. Here we simply note that the term $\textrm{Ei}(-2 \epsilon \Delta\omega)$ generates the familiar logarithmic divergence in the response function in the limit $\epsilon \to 0$. By the definition (\ref{rate1}), the associated transition rate reads
\begin{eqnarray}
\hspace{-5mm}
R_{11}(\Delta \omega ,\Delta t) &=& \frac{1}{2\pi ^{2}}
\left\lbrace e^{-2\epsilon|\Delta \omega|}\pi |\Delta \omega |\Theta (-\Delta \omega)\left[1 + \frac{1}
{e^{2\pi \alpha _{1}|\Delta \omega |}-1}\right] \right.
\nonumber\\
&+&\, \left. \frac{e^{2\epsilon\Delta \omega}\pi\Delta \omega\Theta (\Delta \omega)}{e^{2\pi \alpha _{1}\Delta \omega }-1}
+ {\cal P}_{1}(\Delta\omega, \Delta t, \epsilon) 
+ \Delta t\frac{\partial {\cal P}_{1}(\Delta\omega, \Delta t, \epsilon)}{\partial (\Delta t)} \right.
\nn\\
&+&\, \left. \Re\left[{\cal J}_{1}(\Delta\omega, \Delta t, \epsilon)\right]
+ \Delta t\Re\left[\frac{\partial {\cal J}_{1}(\Delta\omega, \Delta t, \epsilon)}{\partial (\Delta t)}\right]
\right\rbrace
\nonumber \\
&+&\,  \frac{1}{2\pi ^{2}}\left\{\frac{\partial {\cal P}_{2}(\Delta\omega, \Delta t, \epsilon)}{\partial(\Delta t)} 
+ \Re\left[\frac{\partial {\cal J}_{2}(\Delta\omega, \Delta t, \epsilon)}{\partial (\Delta t)}\right]\right\}.
\label{r11sec1}
\end{eqnarray}

\section{Explicit calculation of the crossed response functions}
\label{app2}

\begin{figure}[h]
\begin{center}
\includegraphics[scale=0.68]{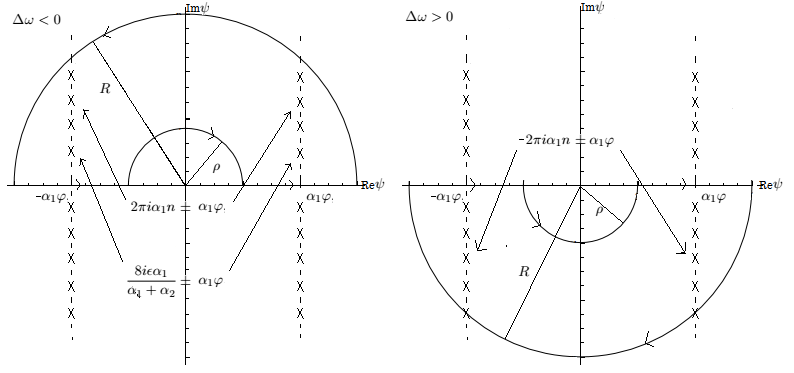}\\
\caption{Contour used for perform the integral of $F_{12}(\Delta \omega ,\Delta t)$ and $F_{21}(\Delta \omega ,\Delta t)$.}
\label{contorno2}
\end{center}
\end{figure}

In this Appendix we perform the explicit evaluation of the cross contributions $F_{12}$ and $F_{21}$ in the case of a sharp switching function. As above, we shall employ the method of residues. The integral (\ref{integral1}) can be expressed as,
\begin{equation}
I_{\epsilon}(\Delta \omega ,\Delta t, \sigma) = \int _{0}^{\infty} d\psi e^{-i\sigma \Delta \omega \psi}G^{+}_{c_{0}}(\psi,\epsilon)- \int _{\Delta t}^{\infty} d\psi e^{-i\sigma \Delta \omega \psi}G^{+}_{c_{0}}(\psi,\epsilon) .
\label{integral2}
\end{equation}
For the first term on the right-hand side of the expression (\ref{integral2}), the simple poles of the integrand are given by
\begin{equation}
\psi^{\pm} _{n}= 2\pi i\alpha _{1}n+\frac{8i\epsilon \alpha _{1}}{\alpha _{1}+\alpha _{2}}\pm \alpha _{1}\phi ,
\label{poles}
\end{equation}
where $n$ is an integer. One should make use of the following auxiliary contour integrals in the complex $z$-plane
\begin{equation*}
\oint _{C}dze^{-i\Delta \omega \sigma z}\log(z)G^{+}_{c_{0}}(z,\epsilon) , \qquad \oint _{C}dze^{-i\Delta \omega \sigma z}
G^{+}_{c_{0}}(z,\epsilon),
\end{equation*}
where $C$ is one of the contours portrayed in Fig.~\ref{contorno2} (the smaller semicircle is introduced here in order to avoid passing through the branch point $z = 0$ of the complex logarithm). In the left (right) figure, the branch cut associated with the logarithm function consists of the origin and the negative (positive) imaginary axis. In turn note that, in order to correctly implement the residue theorem, one should take into account both signs of the quantities $\sigma$ and $\Delta \omega$. When the signs are different and $\epsilon > 0$ ($\epsilon < 0$), one performs the integrals with the contour depicted on the left of the Fig.~\ref{contorno2} such that it encloses the poles for $n\geq 0$ ($n > 0$) and runs in an anticlockwise direction. On the other hand, when both signs are equal and $\epsilon > 0$ ($\epsilon < 0$) one employs the contour on the right such that it encloses the poles for $n < 0$ ($n \leq 0$) and runs in the clockwise direction. Proceeding as described above, the contributions from the larger semicircles will vanish in the limit $R \to \infty$ as ensured by Jordan's lemma. On the other hand, we emphasize that care must be taken in specifying the argument of a complex number since the above contours embraces different branches of the logarithm function: for a complex number $z$, the left contour stipulates that $-\pi/2 < \arg(z) < 3\pi/2$, whereas for the right contour one has that $-3\pi/2 < \arg(z) < \pi/2$, where $\arg(z)$ is the (multi-valued function) argument of $z$.

As an example of such considerations, let us evaluate the improper integral in Eq.~(\ref{integral2}) for the case in which 
$\sigma >0$ and $\Delta\omega < 0$. Based on the foregoing discussion, it is easy to see that, in the limit $\rho \rightarrow 0$ and $R \rightarrow \infty$:
\bea
\oint _{C}dz\,e^{i\sigma|\Delta \omega| z}G^{+}_{c_{0}}(z,\epsilon) &=& 
\int_{0}^{\infty} d\psi\,e^{i\sigma|\Delta \omega| \psi}G^{+}_{c_{0}}(\psi,\epsilon) 
\nn\\
&+&\, \int_{0}^{\infty} d\psi\,e^{-i\sigma|\Delta \omega| \psi}G^{+}_{c_{0}}(\psi,-\epsilon) 
\nn\\
&=& 2\pi i \sum_{\delta = \pm}\sum_{n=0}^{\infty}\Res\left[e^{i\sigma|\Delta \omega| z}G^{+}_{c_{0}}(z,\epsilon);
\, z = \psi^{\delta} _{n}\right]
\eea
and
\bea
\oint _{C}dz\,e^{i\sigma|\Delta \omega| z}\log(z)G^{+}_{c_{0}}(z,\epsilon) 
&-& \oint _{C}dz\,e^{-i\sigma|\Delta \omega| z}\log(z)G^{+}_{c_{0}}(z,-\epsilon) 
\nn\\
&=& \int_{0}^{\infty} d\psi\,e^{-i\sigma|\Delta \omega| \psi}G^{+}_{c_{0}}(\psi,-\epsilon) - 
\int_{0}^{\infty} d\psi\,e^{i\sigma|\Delta \omega| \psi}G^{+}_{c_{0}}(\psi,\epsilon) 
\nn\\
&=& 2\sum_{\delta = \pm}\sum_{n=0}^{\infty}\Res\left[e^{i\sigma|\Delta \omega| z}\log(z)G^{+}_{c_{0}}(z,\epsilon);
\,z = \psi^{\delta} _{n}\right]
\nn\\
&+& 2\sum_{\delta = \pm}\sum_{n=0}^{\infty}\Res\left[e^{-i\sigma|\Delta \omega| z}\log(z)G^{+}_{c_{0}}(z,-\epsilon); 
\,z = \psi^{\delta} _{-n}\right].
\eea
Hence one gets for $\sigma > 0$ and $\Delta\omega < 0$ 
\bea
\hspace{-11mm}
\int_{0}^{\infty} d\psi\,e^{i\sigma|\Delta \omega| \psi}G^{+}_{c_{0}}(\psi,\epsilon) &=& 
\pi i \sum_{\delta = \pm}\sum_{n=0}^{\infty}\Res\left[e^{i\sigma|\Delta \omega| z}G^{+}_{c_{0}}(z,\epsilon);
\, z = \psi^{\delta} _{n}\right]
\nn\\
&-& \sum_{\delta = \pm}\sum_{n=0}^{\infty}\Res\left[e^{i\sigma|\Delta \omega| z}\log(z)G^{+}_{c_{0}}(z,\epsilon);
\,z = \psi^{\delta} _{n}\right]
\nn\\
&-& \sum_{\delta = \pm}\sum_{n=0}^{\infty}\Res\left[e^{-i\sigma|\Delta \omega| z}\log(z)G^{+}_{c_{0}}(z,-\epsilon); 
\,z = \psi^{\delta} _{-n}\right].
\eea
On the other hand, by using the same arguments as above, one gets, for $\sigma > 0$ and $\Delta\omega > 0$ 
\bea
\hspace{-10mm}
\int_{0}^{\infty} d\psi\,e^{-i\sigma\Delta\omega \psi}G^{+}_{c_{0}}(\psi,\epsilon) &=& 
-\pi i \sum_{\delta = \pm}\sum_{n=1}^{\infty}\Res\left[e^{-i\sigma\Delta \omega z}G^{+}_{c_{0}}(z,\epsilon);
\, z = \psi^{\delta} _{-n}\right]
\nn\\
&+& \sum_{\delta = \pm}\sum_{n=1}^{\infty}\Res\left[e^{-i\sigma\Delta \omega z}\log(z)G^{+}_{c_{0}}(z,\epsilon);
\,z = \psi^{\delta} _{-n}\right]
\nn\\
&+& \sum_{\delta = \pm}\sum_{n=1}^{\infty}\Res\left[e^{i\sigma\Delta \omega z}\log(z)G^{+}_{c_{0}}(z,-\epsilon); 
\,z = \psi^{\delta} _{n}\right].
\eea
Therefore collecting the results just derived, one arrives at the following final expression, for $\sigma > 0$:
\begin{eqnarray}
\hspace{-10mm}
\int _{0}^{\infty} d\psi e^{-i \sigma \Delta \omega \psi}G^{+}_{c_{0}}(\psi,\epsilon) &=&
\frac{4 i \alpha_1}{\sinh(\phi)}\Theta(-\Delta\omega)
\Biggl\{\kappa_{1}(|\Delta\omega|,\sigma\alpha_1, -\epsilon)\,
\left(1 + \frac{1}{e^{2 \pi  \alpha _1 \sigma|\Delta\omega|} - 1}\right)
\nn\\
&+&\, e^{-i \alpha _1 \sigma|\Delta\omega| \phi}\,e^{-\frac{8 \alpha_1\sigma \Delta\omega\epsilon}{\alpha _1+\alpha _2}} 
\Im\left[\Phi^{(0,1,0)}\left(e^{-2 \pi  \alpha _1 \sigma|\Delta\omega|},0,\chi(\epsilon)\right)\right]\Biggr\}
\nn\\
&-&\,\frac{4 i \alpha_1}{\sinh(\phi)}\Theta(\Delta\omega)
\Biggl\{\kappa_{2}(\Delta\omega, \sigma\alpha_1, \epsilon)\,
\frac{1}{e^{2 \pi  \alpha _1 \sigma\Delta\omega} - 1}
\nn\\
&-&\,e^{-i \alpha _1 \sigma\Delta\omega \phi}\,e^{-2 \pi  \alpha _1 \sigma\Delta\omega}\,
e^{\frac{8 \alpha _1 \sigma\Delta\omega\epsilon}{\alpha _1+\alpha _2}} 
\Im\left[\Phi^{(0,1,0)}\left(e^{-2 \pi  \alpha _1 \sigma\Delta\omega},0,1+\chi(-\epsilon)\right)\right]\Biggr\},
\end{eqnarray}
where we have defined the quantities
\beq
\kappa_{1}(\Delta\omega, x, \epsilon) = -\frac{\pi e^{\frac{8 x\Delta\omega\epsilon}{\alpha _1+\alpha _2}}
e^{ i x\Delta\omega\phi}}{2},
\eeq
and
\beq
\kappa_{2}(\Delta\omega, x, \epsilon) = \frac{\pi e^{\frac{8 x \Delta\omega\epsilon}{\alpha _1+\alpha _2}} 
e^{-i x \Delta\omega\phi}\,(-2+3 e^{2 i x \Delta\omega \phi })}{2}.
\eeq
In addition, $\Phi \left( z,s,\alpha \right)$ is the Lerch transcendent function~\cite{series1}, $\Phi ^{(0,1,0)}\left(z,s,\alpha \right)$ is its first derivative with respect to its second argument and 
$$
\chi(\epsilon) =\frac{i\phi}{2\pi} + \frac{4\epsilon}{\pi(\alpha _1+\alpha _2)}.
$$
A similar expression holds for the case in which $\sigma < 0$; in this case the parts concerning $\Delta\omega > 0$ and $\Delta\omega < 0$ are reversed. Notice the importance of taking into account the relationship between the signs of $\sigma$ and $\Delta\omega$. 

In the same way, one can show that ($\sigma > 0$)
\begin{eqnarray}
\int _{0}^{\infty} d\psi e^{-i \sigma \Delta \omega \psi}G^{+}_{c_{0}}(\psi,-\epsilon) &=& 
\frac{4 i \alpha_1}{\sinh(\phi)}\Theta(-\Delta\omega)
\Biggl\{\kappa_{1}(|\Delta\omega|, \sigma\alpha_1, \epsilon)\,
\frac{1}{e^{2 \pi  \alpha _1 \sigma|\Delta\omega|} - 1}
\nn\\
&+&\, e^{-i \alpha _1 \sigma|\Delta\omega| \phi}\,e^{-2 \pi  \alpha _1 \sigma|\Delta\omega|} 
e^{\frac{8 \alpha_1\sigma \Delta\omega\epsilon}{\alpha _1+\alpha _2}}
\Im\left[\Phi^{(0,1,0)}\left(e^{-2 \pi  \alpha _1 \sigma|\Delta\omega|},0,1+\chi(-\epsilon)\right)\right]\Biggr\}
\nn\\
&-&\,\frac{4 i \alpha_1 }{\sinh(\phi)}\Theta(\Delta\omega)
\Biggl\{\kappa_{2}(\Delta\omega, \sigma\alpha_1, -\epsilon)\,
\left(1+\frac{1}{e^{2 \pi  \alpha _1 \sigma\Delta\omega} - 1}\right)
\nn\\
&-&\, e^{-i \alpha _1 \sigma\Delta\omega \phi}e^{-\frac{8 \alpha_1\sigma \Delta\omega\epsilon}{\alpha _1+\alpha _2}} 
\Im\left[\Phi^{(0,1,0)}\left(e^{-2 \pi  \alpha _1 \sigma\Delta\omega},0,\chi(\epsilon)\right)\right]\Biggr\}.
\end{eqnarray}
Gathering our results, one finds that
\bea
\hspace{-10mm}
F_{21}(\Delta \omega ,\Delta t) &=& \frac{1}{4\pi ^{2}\sinh(\phi)\Delta \omega a_{-}}
\Biggl\{ e^{-i\Delta \omega (a_{-}/2\alpha _{1})(\Delta t + 2 T)} 
\left[{\cal A}(\Delta\omega, \alpha_1, \alpha_2) + {\cal I}(\Delta\omega, \Delta t)\right]
\nn\\
&+&\, e^{i\Delta \omega (a_{-}/2\alpha _{1})(\Delta t- 2 T)}
\left[-{\cal A}(\Delta\omega, \alpha_2, \alpha_1) + {\cal I}^{*}(\Delta\omega, \Delta t)\right]
\Biggr\},
\eea
where we have defined
\bea
\hspace{-10mm}
{\cal A}(\Delta\omega, \alpha_1, \alpha_2) &=& \Theta(\Delta\omega)\Biggl\{
\kappa_{1}(\Delta\omega,\alpha_2,\epsilon)\,
\frac{1}{e^{2 \pi  \alpha _2\Delta\omega} - 1} 
\nn\\
&+&\, \kappa_{2}(\Delta\omega, \alpha_1, \epsilon)\,\frac{1}{e^{2 \pi  \alpha _1 \Delta\omega} - 1}
+ \Lambda_{1}(\Delta\omega,\alpha_2,\alpha_1,\epsilon)\Biggr\}
\nn\\
&+&\,\Theta(-\Delta\omega)
\Biggl\{\kappa_{1}(|\Delta\omega|, \alpha_1,-\epsilon)\,
\left(1+\frac{1}{e^{2 \pi  \alpha _1|\Delta\omega|} - 1}\right) 
\nn\\
&+& \kappa_{2}(|\Delta\omega|, \alpha_2, -\epsilon)\,
\left(1 + \frac{1}{e^{2 \pi  \alpha _2 |\Delta\omega|} - 1}\right) 
+ \Lambda_{0}(\Delta\omega,\alpha_1,\alpha_2,-\epsilon)\Biggr\} 
\eea
with 
\bea
\hspace{-10mm}
\Lambda_{\delta}(\Delta\omega,x,y,\epsilon) &=& e^{\frac{8 x\Delta\omega\epsilon}{x+y}} e^{-i x\Delta\omega \phi}\,
e^{-2 \pi  x\delta\Delta\omega} 
\Im\left[\Phi^{(0,1,0)}\left(e^{-2 \pi  x\Delta\omega},0,\delta+\chi(-\epsilon)\right)\right] 
\nn\\
&-&\,e^{\frac{8 y \Delta\omega\epsilon}{x+y}}
 e^{-i y \Delta\omega \phi}e^{-2 \pi  y \delta\Delta\omega} 
\Im\left[\Phi^{(0,1,0)}\left(e^{-2 \pi  y \Delta\omega},0,\delta+\chi(-\epsilon)\right)\right]
\eea
and finally
\beq
{\cal I}(\Delta\omega, \Delta t) = -i\frac{\sinh(\phi)}{4\alpha _1}
\left[\int _{\Delta t}^{\infty} d\psi e^{- i\Delta \omega \psi}G^{+}_{c_{0}}(\psi,\epsilon) 
+ \int _{\Delta t}^{\infty} d\psi e^{i(\alpha_{2}/\alpha_{1})\Delta\omega\psi}
G^{+}_{c_{0}}(\psi,-\epsilon)\right].
\eeq
We observe contributions with $\beta ^{-1}_{1}=1/(2\pi \alpha _{1})$ as well as terms related to 
$\beta ^{-1}_{2}=1/(2\pi \alpha _{2})$. 

As explained in the text, since $F_{12} = F^{*}_{21}$, one simply needs the real part of the above expression. Accordingly, regarding the total transition rate, the contribution of the cross correlations is given by twice the real part of the following expression
\bea
R_{21}(\Delta \omega ,\Delta t) &=& \frac{1}{4\pi ^{2}\sinh(\phi)\Delta \omega a_{-}}
\left\{-\frac{i\Delta \omega a_{-}}{\alpha _{1}}\,e^{-i\Delta \omega (a_{-}/2\alpha _{1})(\Delta t + 2 T)} 
\left[{\cal A}(\Delta\omega, \alpha_1, \alpha_2) + {\cal I}(\Delta\omega, \Delta t)\right]
\right. 
\nn\\
&+&\, \left. e^{-i\Delta \omega (a_{-}/2\alpha _{1})(\Delta t+2T)}\,\frac{\partial {\cal I}(\Delta\omega, \Delta t)}{\partial (\Delta t)} \right.
\nn\\
 &+&\, \left. \frac{i\Delta \omega a_{-}}{\alpha _{1}}\,e^{i\Delta \omega (a_{-}/2\alpha _{1})(\Delta t- 2 T)}
\left[-{\cal A}(\Delta\omega, \alpha_2, \alpha_1) + {\cal I}^{*}(\Delta\omega, \Delta t)\right] \right.
\nn\\ 
&+&\, \left. e^{i\Delta \omega (a_{-}/2\alpha _{1})(\Delta t- 2 T)}\,\frac{\partial {\cal I}^{*}(\Delta\omega, \Delta t)}{\partial (\Delta t)}
\right\}.
\label{r12app}
\eea

\end{document}